# The Control of Lung Branching Morphogenesis

Dagmar Iber[1,2,*]

**Affiliations:**

[1] Department of Biosystems, Science and Engineering (D-BSSE), ETH Zurich, Mattenstraße 26, 4058 Basel, Switzerland

[2] Swiss Institute of Bioinformatics (SIB), Mattenstraße 26, 4058 Basel, Switzerland

* Corresponding author: Dagmar Iber dagmar.iber@bsse.ethz.ch

**Branching morphogenesis generates epithelial trees which facilitate gas exchange, filtering, as well as secretion processes with their large surface to volume ratio. In this review, we focus on the developmental mechanisms that control the early stages of lung branching morphogenesis. Lung branching morphogenesis involves the stereotypic, recurrent definition of new branch points, subsequent epithelial budding, and lung tube elongation. We discuss current models and experimental evidence for each of these steps. Finally, we discuss the role of the mesenchyme in determining the organ-specific shape.**

**Introduction**

Reptiles, birds, and mammals all rely on their lungs to breathe. Gas exchange is highly efficient within the limited space of the thoracic cavity because the tree-like airway system greatly enlarges the contact surface between blood vessels and airways (Fig. 1A). The typical human lung tree undergoes 23 generations of dichotomous branching (Weibel and Gomez, 1962). The first 16 branch generations are conducting airways (bronchi and bronchioles) and thus serve a structural function in creating the branched architecture. At the distal end of each conducting airway, an acinus forms. Acini are the functional units of the lung, and are composed of a small tree of gas exchanging airways (respiratory bronchioles, alveolar ducts, and alveolar sacs) built from the remaining 7 branch generations (Schittny, 2017). The resulting gas-exchanging surface area of 40-130 square meters is more than 250-times larger than that of a spherical organ with the same volume (Glenny, 2011; Weibel and Gomez, 1962). The combination of a huge surface area and a very thin air-blood barrier maximises gas exchange. Narrow tubes would minimize the dead volume of the conducting airways, but only at the price of increased resistance to ventilation. The optimal design of the conducting airways tree is a fractal-like architecture, where the branch length and diameter shrink by a constant factor, $2^{-1/3} \approx 0.8$, in each branch generation (Fig. 1B) (Mauroy et al., 2004; Wilson, 1967). The ratio between the branch length and diameter is conserved between branch generations, but differs between species (Fig. 1C) (Nelson et al., 1990; West et al., 1986). In humans, branches are 3-fold longer than wide (Nelson et al., 1990). As a result of dichotomous branching, the particular shrinkage factor, and the conserved relationship between branch length and width, the combined volume of all branches in a branch generation is conserved. All branch volumes together add up to a total of 5-6 liters in humans.

The branching architecture of the human lung is highly conserved, with airway branch variants found in less than 30% of a multi-ethnic population (Smith et al., 2018). The observed airway variants were associated with higher odds of chronic obstructive pulmonary disease (COPD) and chronic bronchitis, and could be linked to single-nucleotide polymorphisms (SNPs) in introns of the gene for Fibroblastic growth factor (FGF)10 (Smith et al., 2018), a key regulatory factor in lung branching morphogenesis (Bellusci et al., 1997; De Moerlooze et al., 2000; Peters et al., 1994; Weaver et al., 2000). Similarly, in mice, the sequence of branching events is highly conserved, with only few deviations between littermates (Blanc et al., 2012; Metzger et al., 2008; Short et al., 2013), suggesting that lung branching morphogenesis is not left to chance. So, how does the lung tree emerge during development?

**Lung Specification**

The lung develops from the ventral anterior foregut, adjacent to the heart (Fig. 2) (Cardoso and Lu, 2006; Herriges and Morrisey, 2014). The particular position where the lung develops along the foregut appears to be set by cardiac mesoderm-derived fibroblastic growth factor (FGF)1 and FGF2, which segregate a common pool of foregut endoderm cells into lung, liver or pancreatic fields in a dose-dependent manner (Serls et al., 2005). Compared to the thyroid and pancreas, the lung field requires particularly high FGF concentrations (Serls et al., 2005). In the absence of cardiac mesoderm or FGF1/FGF2, the endoderm adopts a default pancreatic fate (Deutsch et al., 2001; Rossi et al., 2001; Serls et al., 2005). The specification of the lung field also depends on canonical WNT2/2b and retinoic acid (RA) signaling (Goss et al., 2009; Que et al., 2006). The homeobox transcription factor *Nkx2-1* is the earliest marker of the developing lung field, but the two endodermal lung buds emerge independent of NKX2.1 around embryonic day (E)9.5 (Kimura et al., 1996; Minoo et al., 1999). They subsequently invade the adjacent mesoderm, and elongate to form the primary buds of the left and right lung. Even though FGF10 promotes *Nkx2*-1 and inhibits *Sox2* expression (Que et al., 2007), FGF10 is only required for the formation of the primary buds, but not for the induction of the lung field or the formation of the trachea ; the ERK(extracellular signal-regulated kinases)/MAPK(mitogen-activated protein kinases) pathway is not required for the formation of the primary buds (Boucherat et al., 2014; De Moerlooze et al., 2000; Hubert et al., 2018; Min et al., 1998; Sekine et al., 1999). The receptor FGFR2 is expressed in the foregut endoderm, while *Fgf10* expression is restricted to the adjacent mesoderm. WNT2/2b signaling enhances mesodermal *Fgf10* expression (Goss et al., 2011; Goss et al., 2009), and WNT signalling can also upregulate endodermal *Fgfr2b* expression (Kadzik et al., 2014). Bone Morphogenetic Protein (BMP) signaling counteracts lung bud formation, and conditional removal of the BMP type I receptor genes *Bmpr1a* and *Bmpr1b*, which are normally upregulated in the lung field, results in ectopic primary lung buds that can undergo branching morphogenesis (Domyan et al., 2011; Ikonomou et al., 2020). Additional removal of the gene for the FGFR2 receptor suppresses the formation of ectopic buds. Sonic Hedgehog (SHH) is not required for the formation of the primary buds, but trachea or lung primordia are not observed in *Gli2*/*Gli3* double mutants (Litingtung et al., 1998; Motoyama et al., 1998; Pepicelli et al., 1998). The specification of the lung field, the emergence of the primary buds, and the separation of the trachea from the future esophagus are thus independent developmental processes that are controlled by distinct regulatory networks.

**Trachea Development**

Between E10 and E11.5, the trachea separates from the dorsal foregut, the future esophagus (Fig. 2). The border between the two tissue types is defined by mutual repression between *Nkx2.1*, which is restricted to the ventral foregut endoderm, and *Sox2*, which is expressed in the dorsal foregut endoderm (Que et al., 2007). BMP signaling represses *Sox2* and thereby enables trachea formation (Domyan et al., 2011), while RA signaling induces *Nkx2.1* expression and is required for tracheal–esophageal separation (Que et al., 2006). The separation of trachea and esophagus requires SHH, which acts downstream of *Nkx2.1* and *Sox2* (Ioannides et al., 2010; Litingtung et al., 1998; Minoo et al., 1999; Motoyama et al., 1998; Pepicelli et al., 1998; Que et al., 2007).

The airways are stabilized by cartilage rings and airway smooth muscles that emerge in complementary domains around the airway epithelium and ensure the passage of airflow to the lungs (Hines et al., 2013; Miller et al., 2004; Park et al., 2010; Pepicelli et al., 1998). There are two symmetry breaks, one along the dorsal-ventral axis, giving rise to the distinct domains of smooth muscle and cartilage, and one along the proximal-distal axis in form of the periodic cartilage ring pattern. The dorsal-ventral split is maintained by the mutual inhibition between NKX2.1 and SOX2 (Que et al., 2007). *Nkx2.1* null mice upregulate *Sox2* and form a continuous ring of smooth muscle and no cartilage rings (Minoo et al., 1999; Que et al., 2007; Yuan et al., 2000). Later interference with cartilage or smooth muscle formation alters the spatial domain of the remaining tissue type in the bronchi, but not in the trachea (Hines et al., 2013).

So, how does the spatial restriction of epithelial transcription factors result in mesenchymal patterning? NKX2.1 directly positively regulates expression of *Wnt7b* in the tracheal epithelium (Kuwahara et al., 2020), and WNT ligands from the tracheal epithelium and mesodermal canonical WNT signalling are required for tracheal mesoderm and cartilage formation (Kishimoto et al., 2020; Snowball et al., 2015). Additional factors must promote WNT production as the tracheal mesoderm is specified also independent of *Nkx2.1* (Kishimoto et al., 2020). Together with SOX2, NKX2.1 also upregulates *Shh* expression (Kuwahara et al., 2020). *Sox2* and *Shh* expression are both higher in the dorsal half (Que et al., 2009). SHH is necessary for smooth muscle formation in the dorsal tracheal mesenchyme (Litingtung et al., 1998; Pepicelli et al., 1998), and for the expression of *Sox9* and type II collagen (*Col2a1*) in the ventral tracheal mesenchyme (Miller et al., 2004; Park et al., 2010). Type II collagen and SOX9 are essential for the differentiation of condensed mesenchymal cells into chondrocytes and subsequent formation of cartilage (Barbieri et al., 2003). Even though *Shh* expression is higher in the dorsal epithelium, overexpression of *Shh* does not change the relative cartilage and smooth muscle domains (Sala et al., 2011). Mesenchymal ERK/MAPK signalling further enhances *Sox9* expression, and mesenchymal loss of *Mek* function results in near-complete loss of the C-shaped cartilage rings along the entire length of the upper airways (Boucherat et al., 2014).

So, how do the cartilage rings assume their periodic patterns? *Shh*, its transcriptional targets *Sox9* and type II collagen (*Col2a1*), as well as active mesenchymal ERK show first signs of periodicity around

E13 (Elluru et al., 2009; Hines et al., 2013; Miller et al., 2004; Park et al., 2010; Sala et al., 2011; Turcatel et al., 2013; Yoshida et al., 2020; Young et al., 2020). *Shh* expression becomes periodic only in the ventral, but not in the dorsal epithelium (Sala et al., 2011). Active mesenchymal ERK, *Sox9* and *Col2a1* are restricted to the same subdomains in the ventral mesenchyme because active ERK drives *Sox9* expression, and SOX9 directly regulates *Col2a1* expression (Boucherat et al., 2014; Rockich et al., 2013; Yoshida et al., 2020). It is not clear which FGF ligand/receptor pair drives the periodic activation of mesenchymal MAPK/ERK. One candidate is FGF18, which is expressed in the tracheal mesenchyme, signals via MAPK/ERK, and drives *Sox9* expression (Elluru et al., 2009). However, there is no tracheal phenotype in *Fgf18* null mice (Usui et al., 2004). SHH appears to engage in a positive feedback with FGF10 in that SHH represses *Fgf10* expression in the mesenchyme (Abler et al., 2009; Bellusci et al., 1997; Park et al., 1998), while FGF10 represses *Shh* expression in the tracheal epithelium (Sala et al., 2011), potentially by repressing *Sox2*, which enhances *Shh* expression (Que et al., 2009; Que et al., 2007). *Fgf10* expression is restricted to the ventral mesenchyme, and weak uniform expression becomes first visible at E12.5 (Sala et al., 2011). From E14.5, *Fgf10* expression becomes restricted to the ventral mesenchyme in between the cartilage rings (Sala et al., 2011). SHH and FGF10 can, in principle, both give rise to periodic patterns via a Turing mechanism (Kurics et al., 2014; Menshykau et al., 2012). The regulatory interactions between SHH and FGF10 (Sala et al., 2011) would then explain why cartilage rings become disorganised in *Shh* and *Fgf10* null mice (Miller et al., 2004; Pepicelli et al., 1998; Sala et al., 2011). Consistent with a Turing mechanism, the spacing between cartilage rings increases when trachea explants are cultured with an ERK inhibitor for two days from E13.5 (Young et al., 2020). However, *Shh* expression still becomes periodic in *Fgf10* null mice, if with one day delay (Sala et al., 2011), and epithelial MAPK/ERK is not required for cartilage ring formation (Boucherat et al., 2014). Further redundant regulatory interactions must exist that result in the periodic patterns. These could be provided by other FGF-receptor pairs or by BMP signalling, which, in principle, can result in Turing patterns (Badugu et al., 2012; Kurics et al., 2014), and which controls ERK activity as well as *Sox9* and *Sox2* expression (Boucherat et al., 2014; Domyan et al., 2011; Park et al., 2010).

**Lung Branching Morphogenesis**

Between E10.5 and E16.5, a highly regulated and repetitive combination of branching and elongation generates the arborized airway network, in a process referred to as branching morphogenesis. Three dominating modes of branching generate the airway tree: lateral domain branching, planar bifurcations, and orthogonal bifurcations of consecutive branching events (Metzger et al., 2008). During domain branching, the epithelium of an existing branch bulges out to form a new branch. Bifurcations emerge when the tip splits either in plane with a previous branching event or orthogonal to the previous branching direction. In the mouse, acini form on average after 13-17 branch generations of a total of 27 branch generations, but can also be observed already after 6 branch generations (Madl et al., 2010; Schittny, 2017). The proximal endoderm, which gives rise to the conducting airways with airway neuroendocrine cells, secretory cells, ciliated cells and mucosal cells, is marked by *Sox2* expression, while the distal endoderm, which gives rise to alveolar epithelial cells, is marked by the combined expression of *Sox9* and *Id2* (Herriges and Morrisey, 2014). WNT/β-catenin and BMP signaling are required to establish the distinct proximal-distal cell fates.

While the esophagus forms a multilayered epithelium, the lung epithelium forms a single layer epithelium, which is pseudostratified in the proximal airways and single simplified in the distal airways. In *Nkx2.1* null mice, the lung endoderm is multilayered and fails to branch after formation of the primary buds (Que et al., 2007). In addition to *Shh*, the expression of alpha-integrins and collagen type IV is strongly reduced or absent (Yuan et al., 2000). Interestingly, conditional disruption of *Itgb1* in the lung epithelium, and thus removal of the major isoform of the eight beta integrin subunits that forms 12 of the 24 known integrin a/b heterodimers, results in a multilayered lung epithelium that can no longer branch (Chen and Krasnow, 2012). In the mutant, the mitotic spindle orientation is no longer mostly parallel to the luminal surface, and apical markers are observed also in the basal-most layer. The epithelial architecture is thus important for branching morphogenesis.

In the following, we will discuss the regulatory mechanism that control distinct hallmarks of the branching process: the stereotypic definition of branch points, the coordination of lung growth and branching, the biased elongation of lung tubes, and the shape of lung bud tips.

**Definition of Branch Points**

As the lung tubes are growing out, new buds emerge either along the stalk or by bifurcation of the tip (Fig. 3A). The sequence by which these branch points appear is highly conserved between embryos (Blanc et al., 2012; Metzger et al., 2008; Short et al., 2013). Several models have been put forward to explain how branch points are defined in the growing lung, as reviewed before (Iber and Menshykau, 2013). FGF10 signalling via the ERK/MAPK pathway is central to the control of lung branching morphogenesis as it is both necessary and sufficient to induce a lung branch (Fig. 3B) (Bellusci et al., 1997; Boucherat et al., 2014; De Moerlooze et al., 2000; Peters et al., 1994; Weaver et al., 2000). *Fgf10* is expressed in the submesothelial mesenchyme, while expression of its receptor, FGFR2b, and thereby FGF10 signalling, is restricted to the epithelium. FGF10 signalling induces the expression of Sonic Hedgehog (*Shh*) in the epithelium, which in turn diffuses into the mesenchyme and represses *Fgf10* (Abler et al., 2009; Bellusci et al., 1997; Park et al., 1998). It has been proposed that because of this negative feedback, *Fgf10* expression levels are lower, the smaller the distance between the mesothelium and the epithelium (Bellusci et al., 1997; Hirashima and Iwasa, 2009). As branches grow out, this distance becomes smaller, and the lower *Fgf10* expression in front of the bud could then result in bifurcating outgrowth of branches. According to an alternative model, the epithelium directly recognizes the distance to the mesothelium via the local amplitude or the steepness FGF10 gradient at the epithelial-mesenchyme border (Clement et al., 2012). Finally, it has been proposed that a protein that inhibits branching is secreted from the epithelium and the epithelial geometry results in the observed signalling pattern and points of outgrowth (Gleghorn et al., 2012). This protein would be SHH or TGF-beta (Gleghorn et al., 2012). None of the above mechanisms would be able to explain branching in mutant lung buds where *Fgf10* is expressed throughout the mesenchyme (Volckaert et al., 2013) or in mesenchyme-free explant cultures where recombinant FGF10 is added uniformly (Bellusci et al., 1997; Ohtsuka et al., 2001; Park et al., 1998).

We have shown that the interaction of the FGF10 ligand with its receptor FGFR2b, and the interaction of SHH with its receptor PTCH can both give rise to ligand-receptor based Turing patterns (Fig. 3C) (Kurics et al., 2014; Menshykau et al., 2014; Menshykau et al., 2012). Turing mechanisms enable the deterministic, stereotyped formation of patterns from noisy initial conditions (Turing, 1952). The ligand-receptor based Turing mechanism is based on the well-known Schnakenberg equations, where the two equations describe the dynamics of the receptor (R) and ligand (L),

$$\frac{\partial R}{\partial t} = \Delta R + \gamma(a - R + R^2 L)$$

$$\frac{\partial L}{\partial t} = d\Delta L + \gamma(b - R^2 L).$$

To generate Turing patterns, the ligand must diffuse faster than the receptors ($d > 1$), which is indeed the case because the receptors diffuse more slowly within the cell membrane than the ligands in the extracellular space. A key aspect of the ligand-receptor based Turing mechanism is that ligands and receptors must create multimeric ligand-receptor complexes (Kurics et al., 2014). FGF10 is indeed a dimer, and SHH a multimer, such that one ligand can engage multiple receptors (Ibrahimi et al., 2005; Zeng et al., 2001). The binding of ligand and receptor must result in the removal of the ligand and the upregulation of the receptors on the cell surface, as is the case for SHH/PTCH (Ingham and McMahon, 2001), and likely also for FGF10/FGFR2b as FGF10 upregulates β-catenin signalling, which in turn upregulates *FfgR2b* expression (Estival et al., 1996; Jones et al., 2018; Kadzik et al., 2014). Interestingly, FGF7, which binds to the same receptor as FGF10 but triggers its removal rather than recycling to the membrane, blocks branching morphogenesis (Francavilla et al., 2013). Finally, in the default ligand-receptor based Turing model, receptor and ligands are produced at constant rates, *a* and *b* respectively. But this is not a necessary condition, and feedbacks that modulate the production rate, can lead to additional important properties, as discussed below.

The ligand-receptor based Turing mechanism explains how FGF10 signalling concentrates at newly forming branch tips (Fig. 4D, E), and produces signalling patterns that are consistent with the observed branching patterns in the wildtype and all known relevant mutants (Cellière et al., 2012; Menshykau et al., 2014; Menshykau et al., 2012). In particular, the Turing mechanism reproduces the larger spacing between branch points in hypomorphic mutants that express *Fgf10* at lower levels (Mailleux et al., 2005), and explains how branching patterns can still arise when *Fgf10* is expressed homogenously throughout the lung mesenchyme (Cellière et al., 2012; Menshykau et al., 2014; Menshykau et al., 2012; Volckaert et al., 2013). When put in proximity, lung buds compete for FGF ligand (Miura and Shiota, 2002). This depletion is consistent with a ligand-receptor based Turing mechanism, and particular relevant in kidney branching morphogenesis (Menshykau et al., 2019). We went on to quantitatively evaluate both the ligand-receptor based Turing models and the other candidate models. Given that both FGF10 and SHH are both necessary and sufficient to induce the formation of new branches during lung development (Bellusci et al., 1997; De Moerlooze et al., 2000; Litingtung et al., 1998; Pepicelli et al., 1998; Peters et al., 1994; Weaver et al., 2000), we evaluated the different models by comparing the predicted pattern of ligand-receptor signalling to the measured embryonic growth fields on 3D shapes of developing embryonic lung buds (Fig. 3E) (Menshykau et al., 2014). We found that only the ligand-receptor based Turing models, but none of the other proposed models succeeded in reproducing the observed growth fields. The ligand-receptor based Turing mechanism not only predicts the branch points, but can also guide the outgrowth of branches as observed in embryonic lungs (Menshykau et al., 2014). Here, we solved the Turing mechanism on an embryonic geometry and grew and deformed the epithelium according to the signalling strength predicted by the Turing mechanism, i.e. we applied a velocity field normal to the epithelial boundary whose strength was proportional to the local intensity of the local ligand-receptor complex pattern (Menshykau et al., 2014). The Turing mechanism, but none of the alternative mechanism, succeeded in giving rise to a stereotyped branched structure (Menshykau et al., 2014). We compared the simulation output to the branching patterns observed with lung explants that were cultured for 48-72h, and the simulations with the ligand-receptor-based Turing mechanism recapitulated the observed branching pattern of lung buds very well. Based on all these studies, we conclude that the Turing mechanism, but none of the other proposed mechanisms, quantitatively recapitulates the lung branching patterns both in wildtype and perturbed signalling conditions.

The Turing mechanism on its own has several limitations. For one, many different patterns can emerge with the same parameter set (Fig. 3F) (Murray, 1993), such that branching sequences would vary between embryos. This is not the case for the ligand-receptor based Turing mechanism because the ligands and receptors are expressed in two different domains in the lung, the mesenchyme and the epithelium (Fig. 3B). As a result, the tissue geometry creates a pre-pattern that biases the Turing pattern to the same pattern in all embryos (Fig. 3G) (Menshykau et al., 2014). When ligands and receptors are both expressed in the kidney epithelium, branching morphogenesis still proceeds, but branching patterns indeed differ between embryos (Shakya et al., 2005).

A second problem concerns the evolution of Turing mechanisms. The parameter space that supports Turing patterns, the so-called Turing space, is small for classical Turing mechanism (Fig. 3H) (Murray, 1982). This would make it unlikely for nature to hit on the relevant parameter space and to retain it during evolution. The confinement of receptors to single cell surfaces greatly enlarges the Turing space (Fig. 3H) (Kurics et al., 2014). Receptor clustering (Fig. 3I) and regulatory interactions between Turing modules (Fig. 3J,K), as they exist between SHH and FGF10, further enlarge the Turing space such that the evolution of this patterning mechanism is highly plausible.

As the lung grows, new branch points emerge at a set distance from previous branch points in Turing models. Turing mechanisms thus provide a direct link between growth and branching. Indeed, faster growth is typically accompanied by enhanced branching and vice versa (Table 1). Enhanced branching can also result from a smaller distance between branch points, which can come about from changes in the reaction or diffusion rates. Many perturbations that affect the growth rate also affect these Turing parameters. Thus, inactivation of any of the key genes for lung branching morphogenesis, *Shh, Fgf10*, and *FgfR2b* impairs embryonic growth (Chiang et al., 1996; De

Moerlooze et al., 2000; Min et al., 1998; Sekine et al., 1999), and the FGF receptor inhibitor SU5402 and Smoothened agonist (SAG) reduce both growth and branching in lung explants (Goodwin et al., 2019; Runser et al., 2020). Conditional removal of the WNT receptor Frizzled in the lung epithelium reduces the growth rate and branching, even though *Fgf10* expression is increased in the mesenchyme (Kadzik et al., 2014). This is likely due to the reduction in *FgfR2* and *Shh* expression in the lung epithelium. WNT/β-catenin signaling enhances *Sox9* expression (Ustiyan et al., 2016), and both, upregulation and inactivation of epithelial *Sox9*, results in smaller lungs with reduced branching (Chang et al., 2013; Rockich et al., 2013).

The growth speed can be increased also mechanically by an increase in luminal pressure, as generated either by tracheal occlusion or with a microfluidic chamber (Nelson et al., 2017; Unbekandt et al., 2008). Increased growth and branching, however, are still linked to increased *Fgf10* expression and are not observed in cauterized lung explants from *FgfR2b* null mice (Unbekandt et al., 2008). Removal of *Adamts18*, which encodes a secreted matrix metalloproteinase (MMP) that degrades the extracellular matrix (ECM), results in shorter primary airways and a proportional reduction in branching without affecting the expression of *Shh*, *Bmp4*, or *Fgf10* (Rutledge et al., 2019). Interestingly, heterozygous *Adamts18* mutants have larger lungs with more branches. Disrupting the actomyosin cytoskeleton or the basement membrane impairs both growth and branching (Moore et al., 2005; Relan and Schuger, 1999).

There are also examples where growth and branching are not linked. In all those cases, the bud tips, and sometimes also the tubes, are dilated. Thus, E11.5 lung explants that are treated with EGF or TGF become larger, but have fewer branches with dilated buds (Ganser et al., 1991). The dilation and branching defect can be prevented by co-culture with Tissue inhibitors of metalloproteinases (TIMPs), which regulate ECM degradation by MMPs. Likewise, addition of collagenase inhibits branching and dilates tips. Similarly, E12 lungs in *Erk* null mice have the same size, but fewer branches with dilated tips (Miettinen et al., 1997). Expression of a hyperactive KRas allele in the lung epithelium results in larger lungs with wider tubes and cystic tips, but unchanged number of branches

(Chang et al., 2013), though it is unclear whether the length of branches is increased compared to the wildtype, or only their width (Tang et al., 2011). In all these cases, branching appears to be impaired because of changes in the epithelial structure and the ECM. This shows that in addition to the chemical definition of new branch points, the structure of the epithelium and the ECM must be such that bud formation is possible, as will be discussed in more detail below.

The potential of Turing mechanisms to explain lung branching morphogenesis was recognised early on, and several other Turing mechanisms have been proposed to explain lung branching morphogenesis (Guo et al., 2014; Meinhardt, 1976; Warburton et al., 2003; Xu et al., 2017; Zhu and Yang, 2018). Together, these mathematical studies demonstrate that important hallmarks of branching morphogenesis can be recapitulated by Turing mechanisms. To test a Turing mechanism, it is important to identify the Turing components and check that the mechanism is consistent with all reported genetic and pharmacological perturbations. To date, this has only been done with the ligand-receptor-based Turing mechanism for the lung (Cellière et al., 2012; Menshykau et al., 2014; Menshykau et al., 2012). Even though the qualitative and quantitative explanation of a wide range of observations provides important support, pattern likeness alone is not sufficient proof that branching morphogenesis is controlled by a Turing mechanism. Ultimately, all parameter values would have to be measured to prove the existence of a Turing pattern, but this is so far impossible in biological tissue.

In addition to signalling mechanisms, mechanical mechanisms have been proposed (Kim et al., 2015; Lubkin and Murray, 1995; Wan et al., 2008). However, these would result in random, rather than stereotyped branching patterns. Finally, differential proliferation in the epithelial and mesenchymal layers has been suggested to result in budding via epithelial buckling (Varner et al., 2015). The distance between buds would then depend on the thickness and the mechanical properties of the layers. The distance between buds in the mesenchyme-free epithelial cultures was indeed found to depend both on the Matrigel concentration and on the FGF1 concentration. In terms of the ligand-receptor based Turing mechanism, this observation could, however, also be explained with a change in the interaction of FGF1 with the extracellular matrix (ECM) and its diffusivity at different Matrigel concentrations (Makarenkova et al., 2009; Miura and Shiota, 2002). The reported shorter distance between branch points at higher FGF1 concentrations is consistent with the ligand-receptor based Turing mechanism.

**Formation of a Bud**

Once the position of a new branch point has been specified, how does the epithelial bud form? Since buds can form also in mesenchyme-free cultures, the epithelium itself must be able to generate buds. Even though growth factor signaling concentrates at the branch point (Fig. 3E) (Liu et al., 2004; Runser et al., 2020), a localized increase in the proliferation rate is not required for bud formation (Kim et al., 2013; Mollard and Dziadek, 1998; Nogawa et al., 1998). The first sign of a nascent bud is an epithelial thickening that results from the lengthening of the cells along the apical–basal axis (Fig. 4A) (Fumoto et al., 2017; Kadzik et al., 2014). The epithelial sheet then bends and creates a bud. Given the geometry of the buds, the apical surface area is smaller than the basal surface area (Fumoto et al., 2017; Kadzik et al., 2014; Kim et al., 2013). Such cell shapes are characteristic for apical constriction (Martin and Goldstein, 2014), and simulations demonstrate that apical constriction in combination with proliferation can generate buds (Fumoto et al., 2017; Kim et al., 2013). The mechanistic details of apical constriction and cell lengthening are still under debate, but the molecular motor non-muscle myosin II has long been implicated in generating the force that drives apical constriction by contracting actin filament (F-actin) networks that are linked to the apical cell-cell adhesion belt via α-catenin and E-cadherin, as well as other protein complexes (Martin and Goldstein, 2014). The extracellular domain of E-cadherin mediates cell-cell adhesion.

Myosin motor activity is regulated by phosphorylation of the myosin regulatory light chain (MLC) (Fig. 4B), and both actin and phosphorylated myosin light chain (pMLC) have been reported to accumulate at the apical surface of lung buds (Fumoto et al., 2017; Kadzik et al., 2014; Kim et al.,

2013; Rockich et al., 2013). Conditional removal of the WNT receptor Frizzled from the lung epithelium interferes with cell lengthening and concomitant epithelial thickening and bud formation (Kadzik et al., 2014). The effects of the Frizzled mutation can be counteracted with calpeptin, a RhoA activator, which activates Rho kinase (ROCK), which in turn phosphorylates MLC, inhibits the myosin phosphatase, and blocks the depolymerisation of actin filaments. Inhibitors of actin and myosin activation impair not only branching, but also growth (Kadzik et al., 2014; Moore et al., 2005).

Ectopic expression of E-cadherin impairs budding of E11.5 mesenchyme free lung epithelium (Liu et al., 2008). Fibronectin supports lung branching morphogenesis by inducing *Btbd7*, which regulates the expression of *Snail2* and E-cadherin, and more generally epithelial motility (Onodera et al., 2010; Sakai et al., 2003). Fibronectin and BTBD7 induce labile cell-cell adhesion and loss of E-cadherin from cell boundaries. Blocking FGF10 signalling or disrupting the downstream mitogen-activated protein kinase (MAPK) p38α increases expression of E-cadherin (Jones et al., 2018; Liu et al., 2008). Deletion of *Mek1/2* in the ureteric bud epithelium leads to accumulation of E-cadherin also on the baso-lateral cell surfaces and a failure to form new buds, though branches continue to elongate (Ihermann-Hella et al., 2014). Branching is impaired also in the lung, but the impact on E-cadherin has not been documented (Boucherat et al., 2014). The β-catenin agonist CHIR enhances apical constriction and budding in E11.5 mesenchyme-free epithelial lungs cultured at low FGF10 concentrations (Fumoto et al., 2017). Biochemical studies suggest that E-cadherin can bind both α-catenin and β-catenin, but not at the same time (Martin and Goldstein, 2014). WNT signalling thus impacts on bud formation both via canonical (β-catenin) and non-canonical (Rho pathway) signalling.

The wrapping of airway smooth muscles has been suggested as additional driver of bud formation (Goodwin et al., 2019), but mutant studies show that airway smooth muscles are dispensable for branching morphogenesis (Young et al., 2020), and buds emerge already long before airway smooth muscles first emerge at E11.5 (Hines et al., 2013; Tang et al., 2011).

**Epithelial Tube Elongation**

Once formed, epithelial tubes initially preferentially elongate (Fig. 5A), and only later switch to isotropic outgrowth, first in the trachea (from E12.5), and later in the most proximal bronchus (from E13.5) (Kishimoto et al., 2018; Runser et al., 2020; Tang et al., 2011; Tang et al., 2018). While proliferation is elevated in the tips of extending branches (Kim et al., 2013; Nogawa et al., 1998), biased outgrowth is not due to tip elongation. Rather, the proliferating stalks show biased elongation (Tang et al., 2011). Cell shape and cell division are biased in the direction of lung outgrowth (Fig. 5B) (Tang et al., 2011; Tang et al., 2018; Yates et al., 2010). Cells have long been known to divide through their mass point and perpendicular to their longest axis, an observation termed Hertwig's rule, or long-axis rule (Fig. 5C) (Hertwig, 1884). In the lung, the mitotic spindle becomes biased once the cell aspect ratio at interphase is higher than 1.53 (Tang et al., 2018). Computational modelling shows that such a bias in cell shape and cell division can, in principle, result in biased outgrowth (Runser et al., 2020; Stopka et al., 2019; Tang et al., 2011). So, what leads to this bias in cell shape?

The planar cell polarity (PCP) pathway generates polarised patterns in epithelial fields and plays an important role in regulating the mitotic spindle angle distribution in other organs like the kidney (Ciruna et al., 2006; Gong et al., 2004; Saburi et al., 2008). However, airway shape and spindle orientation are normal in E11.5 lungs from Van Gogh-like (*Vangl)2*$^{Lp}$ homozygotes (Tang et al., 2011), though the number of epithelial buds is lower and the bud width is wider after 48 hours of culture (Yates et al., 2010). Independent of whether the PCP pathway is involved, an additional mechanism would still be required to set the direction of polarisation along the tube. In principle, such a bias could originate from a polarization along the tube, from a mechanical constraint that limits expansion in the circumferential direction, or from a pulling force at the tip or inside the tube in form of flow-induced shear stress (Fig. 5D).

While FGF10 can serve as chemoattractant for explant lungs (Fig. 5E) (Park et al., 1998), branching morphogenesis and biased tube elongation are still observed when FGF10 is added homogenously to explant cultures (Bellusci et al., 1997; Ohtsuka et al., 2001; Park et al., 1998; Runser et al., 2020). Moreover, lung epithelial tubes continue to elongate in *Fgf10* and *Fgfr2* conditional mutants and when an FGF receptor inhibitor is added homogenously to explant cultures (Abler et al., 2009; Runser et al., 2020). Accordingly, neither FGF gradients nor focused FGF signalling at the epithelial tips are required for biased epithelial outgrowth (Fig. 5E). There is also no evidence for epithelial leader cells with actin-rich protrusions (Fig. 5F), as previously observed in the *Drosophila* trachea and the mammalian kidney (Chi et al., 2009; Ghabrial and Krasnow, 2006). Also a pulling force from the tip mesenchyme is not necessary as the mesenchyme is not required for biased epithelial outgrowth (Runser et al., 2020).

Circumferential mechanical constraints have been proposed to drive epithelial tube elongation in the mammary gland (Fig. 5G) (Paine et al., 2016). Lungs do not have a myoepithelium like mammary glands, but are lined by airway smooth muscles, which result in peristaltic fluid movement from E12.5 (Jesudason et al., 2006). However, lung branches are thinner rather than wider when inactivation of *Myocardin* prevents the formation of airway smooth muscles (Young et al., 2020), and biased lung tube elongation is strongest before airway smooth muscles become detectable at E11.5 (Hines et al., 2013; Tang et al., 2011), and is observed also in the absence of mesenchyme (Runser et al., 2020). Furthermore, cell-based simulations show that constricting forces that generate the observed bias in outgrowth, result in a much lower cell shape bias than what is observed in the embryonic lung epithelium (Stopka et al., 2019). Constricting forces are therefore unlikely to drive the elongating outgrowth of lung tubes.

MMPs degrade ECM throughout embryogenesis and thereby enable the invasion of the epithelial bud into the mesenchyme. The basement membrane, composed of nidogen, collagen, fibronectin, and laminin, is thinner at the lung tips (Fig. 5H) (Mollard and Dziadek, 1998; Moore et al., 2005), and epithelial tube elongation sensitively depends on the level of ECM turn-over. Thus, removal of the secreted MMP *Adamts18* enhances epithelial growth in heterozygous mutants and lowers growth in homozygous mutants (Rutledge et al., 2019). *Timp3* mutants have lower fibronectin and E-cadherin levels, lower epithelial proliferation rates, and fewer branches in the lung (Gill et al., 2003; Gill et al., 2006). Low concentrations of the MMP inhibitor GM60001 can rescue the *Timp3* phenotype, while high concentrations completely block MMP activity, preventing the digestion of the ECM ahead of the lung bud, and therefore invasion of the bud into the mesenchyme. It is currently not known to what extent the ECM biases the outgrowth of epithelial tubes, but it would have to do so without exerting a constricting force.

Finally, fluid flow could bias the direction of outgrowth (Fig. 5I). A tangential apical force, as would result from shear stress, can indeed result in both the reported bias in cell shape and elongating outgrowth in cell-based simulations (Runser et al., 2020). Flow-induced shear stress does not deform cells directly, but cells sense shear stress with their cilium and respond actively (Weinbaum et al., 2011). For the observed narrow luminal space (Runser et al., 2020), the reported flow rates ($\dot{V} = 420\ \mu m^3/s$ (George et al., 2015), and the 10-times higher viscosity of the luminal fluid compared to water (Bokka et al., 2015), the expected shear stress level (0.5-1 Pa) in the lung buds is within the range that cells can, in principle, sense ($10^{-3}$ - 5 Pa) (Nauli et al., 2003)(Flitney et al., 2009; Molladavoodi et al., 2017; Resnick and Hopfer, 2007; Weinbaum et al., 2011). The primary cilium, which is required to sense shear stress, is lost in KRas$^{G12D}$ mutants (Tape et al., 2016). Consistent with a role of shear stress in driving elongating outgrowth, the bias in cell division and epithelial outgrowth is lost in embryonic lungs that express the constitutively active form KRas$^{G12D}$ in the epithelium (Shh$^{cre/+}$;KRas$^{LSL-G12D/+}$) (Tang et al., 2011).

In later developmental stages, the closed epithelial lung tubes open up (Runser et al., 2020). Given the larger tube diameter, shear stress levels are expected to drop below a level that cells can sense. Consistent with this, tube outgrowth becomes isotropic in the trachea from E12.5, and in the left bronchus from E13.5 (Kishimoto et al., 2018; Runser et al., 2020). Blockage of FGF10 signalling results in the collapse of bronchi, but how this affects tube outgrowth has not been studied (Jones et al., 2018). Similarly, fluid flow in the lung tubes has so far only been studied from E12.5, while biased outgrowth is observed between E10.5 and E13.5 in the left bronchus. It will be important to measure flow rates also in the earlier stages of lung development, and to understand how cell signalling, fluid dynamics, and mechanical effects integrate to define airway shape.

**Lung Bud Shape**

A wide range of mutations and pharmacological reagents affect the shape of the distal lung tips (Table 1). Most notably, addition of FGF7 or FGF9 to lung explants results in inflated epithelial tips, while lung explants that are treated with the FGF receptor inhibitor SU5402 have narrower tips (Cardoso et al., 1997; del Moral et al., 2006; Runser et al., 2020; White et al., 2006). The epithelium in the inflated FGF9-treated tips tends to be thinner (del Moral et al., 2006). We have seen above that EGF, TGF, hyperactive KRas, and loss of *Erk* all result in dilated tips, and that Tissue inhibitors of metalloproteinases (TIMPs), which regulate ECM degradation by MMPs, can counteract the TGF effect (Ganser et al., 1991; Miettinen et al., 1997) (Tang et al., 2011). Hyperactive KRas inhibits WNT/β-catenin signaling, and WNT/β-catenin signaling upregulates FGFR2b in the lung epithelium (Chang et al., 2013; Kadzik et al., 2014). Hyperactive KRas, WNT/β-catenin signaling, and FGFR2b all support *Sox9* expression (Chang et al., 2013; Kadzik et al., 2014; Ustiyan et al., 2016). Conditional removal of the WNT receptor Frizzled, and both upregulation and removal of *Sox9* expression in the lung epithelium result in larger, rounder tips with open, cyst-like structures (Kadzik et al., 2014; Rockich et al., 2013). SOX9 affects mainly the cytoskeleton and the extracellular matrix (Rockich et al., 2013). In scratch assays, epithelial cells with disrupted *Sox9* expression are less motile. While actin remains distributed normally inside the epithelial cells, there is a reduction in cell-cell adhesion and stabilized tubulin becomes disrupted on the basal side of *Sox9* negative epithelia. Moreover, laminin deposition becomes affected when *Sox9* is removed or overexpressed, and laminin affects the cell shape and polarity in the lung epithelium (Chang et al., 2013) (Nguyen and Senior, 2006; Schuger et al., 1997). SOX9 directly regulates transcription of Type 2 collagen (*Col2a1*), and treatment of lung explants with collagenase results in cyst-like shapes (Chang et al., 2013; Miura and Shiota, 2002). Buds are also widened when lung explants are cultured with cytochalasin D, a drug that disrupts actin microfilament integrity, fasudil, which inhibits Rho kinase (ROCK), or ML7 or ML9, which inhibits myosin light chain kinase (Kadzik et al., 2014; Moore et al., 2005). The tip shape thus depends on the mechanical state of both the epithelial cells and the extracellular matrix.

**The Mesenchyme defines the Shape of the Epithelial Tree**

The shape of epithelial trees differs profoundly between organs. The shape of a branched tree is determined by the distance between branch points, the branch angle, curvature, and aspect ratio. Recombination experiments demonstrate that the mesenchyme strongly affects these parameters. Thus, the epithelial tree that emerges when a ureteric bud is cultured with lung mesenchyme resembles more a lung epithelial tree than that of a kidney (Lin et al., 2003). The growth factors FGF10 and GDNF, that define the branch points in the lung and kidney respectively, are expressed in the lung and kidney mesenchyme (Bellusci et al., 1997; Moore et al., 1996; Pepicelli et al., 1997; Pichel et al., 1996), but beyond this, the mesenchymal factors that determine organ shape are largely elusive. Interestingly, only the bronchial mesenchyme can support lung branching morphogenesis. Mesenchyme from other sources, including the salivary gland, kidney, pancreas, stomach, intestine, dermis, and somites, can induce the formation of a single bud, but cannot support several rounds of branching (Spooner and Wessells, 1970), unless it is renewed (Lawson, 1983). In the lung, the expansion of the mesenchyme and epithelium may be linked via an epithelial-mesenchymal feed-back between FGF and SHH signalling. FGF9 and SHH both increase mesenchymal proliferation though in different parts of the mesenchyme, and the FGF receptor inhibitor SU5402 reduces mesenchymal proliferation (White et al., 2006). Several other morphogens promote epithelial and/or mesenchymal proliferation, including BMP and canonical WNT signalling (Volckaert and De Langhe, 2015).

## Concluding Remarks

There are still many open questions regarding lung branching morphogenesis. Are branch points really defined by a ligand-receptor based Turing mechanism? How does localized FGF10/ERK signalling induce bud formation? When do epithelial cells start to secrete fluid into the lung lumen and thereby create a flow that cells can sense? Are there other mechanisms that bias the outgrowth of lung tubes? What role does the ECM play, and how does the mesenchyme define the specific organ shapes? And how does the final fractal lung tree emerge during development? Unlike the final lung tree, the developing lung tree does not have a fractal architecture. As imaging, genetic and computational techniques become ever more powerful, these questions are now in reach to be answered.

## Acknowledgments

I thank members of the CoBi group for discussions.

# REFERENCES

**Abler, L. L., Mansour, S. L. and Sun, X.** (2009). Conditional gene inactivation reveals roles for Fgf10 and Fgfr2 in establishing a normal pattern of epithelial branching in the mouse lung. *Dev Dyn* **238**, 1999-2013.

**Ali, M.** (2010). Pulmonary Drug Delivery. In *Handbook of Non-Invasive Drug Delivery Systems* (ed. V. S. Kulkarni), pp. 209-246: Elsevier.

**Badugu, A., Kraemer, C., Germann, P., Menshykau, D. and Iber, D.** (2012). Digit patterning during limb development as a result of the BMP-receptor interaction. *Sci Rep* **2**, 991.

**Barbieri, O., Astigiano, S., Morini, M., Tavella, S., Schito, A., Corsi, A., Di Martino, D., Bianco, P., Cancedda, R. and Garofalo, S.** (2003). Depletion of cartilage collagen fibrils in mice carrying a dominant negative Col2a1 transgene affects chondrocyte differentiation. *American journal of physiology. Cell physiology* **285**, C1504-1512.

**Bellusci, S., Grindley, J., Emoto, H., Itoh, N. and Hogan, B. L.** (1997). Fibroblast growth factor 10 (FGF10) and branching morphogenesis in the embryonic mouse lung. *Development* **124**, 4867-4878.

**Blanc, P., Coste, K., Pouchin, P., Azaïs, J.-M., Blanchon, L., Gallot, D. and Sapin, V.** (2012). A role for mesenchyme dynamics in mouse lung branching morphogenesis. *PLoS ONE* **7**, e41643.

**Bokka, K. K., Jesudason, E. C., Lozoya, O. A., Guilak, F., Warburton, D. and Lubkin, S. R.** (2015). Morphogenetic Implications of Peristalsis-Driven Fluid Flow in the Embryonic Lung. *PLoS One* **10**, e0132015.

**Boucherat, O., Nadeau, V., Berube-Simard, F. A., Charron, J. and Jeannotte, L.** (2014). Crucial requirement of ERK/MAPK signaling in respiratory tract development. *Development* **141**, 3197-3211.

**Cardoso, W. V., Itoh, A., Nogawa, H., Mason, I. and Brody, J. S.** (1997). FGF-1 and FGF-7 induce distinct patterns of growth and differentiation in embryonic lung epithelium. *Developmental Dynamics* **208**, 398-405.

**Cardoso, W. V. and Lu, J.** (2006). Regulation of early lung morphogenesis: questions, facts and controversies. *Development* **133**, 1611-1624.

**Cellière, G., Menshykau, D. and Iber, D.** (2012). Simulations demonstrate a simple network to be sufficient to control branch point selection, smooth muscle and vasculature formation during lung branching morphogenesis. *Biol Open* **1**, 775-788.

**Chang, D. R., Martinez Alanis, D., Miller, R. K., Ji, H., Akiyama, H., McCrea, P. D. and Chen, J.** (2013). Lung epithelial branching program antagonizes alveolar differentiation. *Proc Natl Acad Sci U S A* **110**, 18042-18051.

**Chen, J. and Krasnow, M. A.** (2012). Integrin Beta 1 suppresses multilayering of a simple epithelium. *PLoS One* **7**, e52886.

**Chi, X., Michos, O., Shakya, R., Riccio, P., Enomoto, H., Licht, J. D., Asai, N., Takahashi, M., Ohgami, N., Kato, M., et al.** (2009). Ret-dependent cell rearrangements in the Wolffian duct epithelium initiate ureteric bud morphogenesis. *Dev Cell* **17**, 199-209.

**Chiang, C., Litingtung, Y., Lee, E., Young, K. E., Corden, J. L., Westphal, H. and Beachy, P. A.** (1996). Cyclopia and defective axial patterning in mice lacking Sonic hedgehog gene function. *Nature* **383**, 407-413.

**Ciruna, B., Jenny, A., Lee, D., Mlodzik, M. and Schier, A. F.** (2006). Planar cell polarity signalling couples cell division and morphogenesis during neurulation. *Nature* **439**, 220-224.

**Clement, R., Douady, S. and Mauroy, B.** (2012). Branching geometry induced by lung self-regulated growth. *Phys Biol* **9**, 066006.

**De Moerlooze, L., Spencer-Dene, B., Revest, J. M., Hajihosseini, M., Rosewell, I. and Dickson, C.** (2000). An important role for the IIIb isoform of fibroblast growth factor receptor 2 (FGFR2) in mesenchymal-epithelial signalling during mouse organogenesis. *Development* **127**, 483-492.

**del Moral, P. M., De Langhe, S. P., Sala, F. G., Veltmaat, J. M., Tefft, D., Wang, K., Warburton, D. and Bellusci, S.** (2006). Differential role of FGF9 on epithelium and mesenchyme in mouse embryonic lung. *Dev Biol* **293**, 77-89.

**Deutsch, G., Jung, J. N., Zheng, M. H., Lora, J. and Zaret, K. S.** (2001). A bipotential precursor population for pancreas and liver within the embryonic endoderm. *Development* **128**, 871-881.

**Domyan, E. T., Ferretti, E., Throckmorton, K., Mishina, Y., Nicolis, S. K. and Sun, X.** (2011). Signaling through BMP receptors promotes respiratory identity in the foregut via repression of Sox2. *Development* **138**, 971-981.

**Elluru, R. G., Thompson, F. and Reece, A.** (2009). Fibroblast growth factor 18 gives growth and directional cues to airway cartilage. *Laryngoscope* **119**, 1153-1165.

**Estival, A., Monzat, V., Miquel, K., Gaubert, F., Hollande, E., Korc, M., Vaysse, N. and Clemente, F.** (1996). Differential regulation of fibroblast growth factor (FGF) receptor-1 mRNA and protein by two molecular forms of basic FGF - Modulation of FGFR-1 mRNA stability. *Journal of Biological Chemistry* **271**, 5663-5670.

**Flitney, E. W., Kuczmarski, E. R., Adam, S. A. and Goldman, R. D.** (2009). Insights into the mechanical properties of epithelial cells: the effects of shear stress on the assembly and remodeling of keratin intermediate filaments. *FASEB J* **23**, 2110-2119.

**Francavilla, C., Rigbolt, K. T., Emdal, K. B., Carraro, G., Vernet, E., Bekker-Jensen, D. B., Streicher, W., Wikstrom, M., Sundstrom, M., Bellusci, S., et al.** (2013). Functional proteomics defines the molecular switch underlying FGF receptor trafficking and cellular outputs. *Mol Cell* **51**, 707-722.

**Fumoto, K., Takigawa-Imamura, H., Sumiyama, K., Kaneiwa, T. and Kikuchi, A.** (2017). Modulation of apical constriction by Wnt signaling is required for lung epithelial shape transition. *Development* **144**, 151-162.

**Ganser, G. L., Stricklin, G. P. and Matrisian, L. M.** (1991). EGF and TGF alpha influence in vitro lung development by the induction of matrix-degrading metalloproteinases. *Int J Dev Biol* **35**, 453-461.

**George, U. Z., Bokka, K. K., Warburton, D. and Lubkin, S. R.** (2015). Quantifying stretch and secretion in the embryonic lung: Implications for morphogenesis. *Mech Dev* **138 Pt 3**, 356-363.

**Ghabrial, A. S. and Krasnow, M. A.** (2006). Social interactions among epithelial cells during tracheal branching morphogenesis. *Nature* **441**, 746-749.

**Gill, S. E., Pape, M. C., Khokha, R., Watson, A. J. and Leco, K. J.** (2003). A null mutation for Tissue Inhibitor of Metalloproteinases-3 (Timp-3) impairs murine bronchiole branching morphogenesis. *Developmental Biology* **261**, 313-323.

**Gill, S. E., Pape, M. C. and Leco, K. J.** (2006). Tissue inhibitor of metalloproteinases 3 regulates extracellular matrix--cell signaling during bronchiole branching morphogenesis. *Dev Biol* **298**, 540-554.

**Gleghorn, J. P., Kwak, J., Pavlovich, A. L. and Nelson, C. M.** (2012). Inhibitory morphogens and monopodial branching of the embryonic chicken lung. *Dev Dyn* **241**, 852-862.

**Glenny, R. W.** (2011). Emergence of matched airway and vascular trees from fractal rules. *J Appl Physiol* **110**, 1119-1129.

**Gong, Y., Mo, C. and Fraser, S. E.** (2004). Planar cell polarity signalling controls cell division orientation during zebrafish gastrulation. *Nature* **430**, 689-693.

**Goodwin, K., Mao, S., Guyomar, T., Miller, E., Radisky, D. C., Kosmrlj, A. and Nelson, C. M.** (2019). Smooth muscle differentiation shapes domain branches during mouse lung development. *Development* **146**.

**Goss, A. M., Tian, Y., Cheng, L., Yang, J., Zhou, D., Cohen, E. D. and Morrisey, E. E.** (2011). Wnt2 signaling is necessary and sufficient to activate the airway smooth muscle program in the lung by regulating myocardin/Mrtf-B and Fgf10 expression. *Dev Biol* **356**, 541-552.

**Goss, A. M., Tian, Y., Tsukiyama, T., Cohen, E. D., Zhou, D., Lu, M. M., Yamaguchi, T. P. and Morrisey, E. E.** (2009). Wnt2/2b and beta-catenin signaling are necessary and sufficient to specify lung progenitors in the foregut. *Dev Cell* **17**, 290-298.

**Guo, Y., Chen, T. H., Zeng, X., Warburton, D., Bostrom, K. I., Ho, C. M., Zhao, X. and Garfinkel, A.** (2014). Branching patterns emerge in a mathematical model of the dynamics of lung development. *The Journal of physiology* **592**, 313-324.

**Haefeli-Bleuer, B. and Weibel, E. R.** (1988). Morphometry of the human pulmonary acinus. *Anat Rec* **220**, 401-414.

**Herriges, M. and Morrisey, E. E.** (2014). Lung development: orchestrating the generation and regeneration of a complex organ. *Development* **141**, 502-513.

**Hertwig, O.** (1884). Das Problem der Befruchtung und der Isotropie des Eies. Eine Theorie der Vererbung. *Jenaische Zeitschrift fur Naturwissenschaft.* **18**, 274.

**Hines, E. A., Jones, M. K., Verheyden, J. M., Harvey, J. F. and Sun, X.** (2013). Establishment of smooth muscle and cartilage juxtaposition in the developing mouse upper airways. *Proc Natl Acad Sci U S A* **110**, 19444-19449.

**Hirashima, T. and Iwasa, Y.** (2009). Mechanisms for split localization of Fgf10 expression in early lung development. *Developmental Dynamics* **238**, 2813-2822.

**Hubert, F., Payan, S. M. and Rochais, F.** (2018). FGF10 Signaling in Heart Development, Homeostasis, Disease and Repair. *Frontiers in genetics* **9**, 599.

**Iber, D. and Menshykau, D.** (2013). The control of branching morphogenesis. *Open biology* **3**, 130088-130088.

**Ibrahimi, O. A., Yeh, B. K., Eliseenkova, A. V., Zhang, F., Olsen, S. K., Igarashi, M., Aaronson, S. A., Linhardt, R. J. and Mohammadi, M.** (2005). Analysis of mutations in fibroblast growth factor (FGF) and a pathogenic mutation in FGF receptor (FGFR) provides direct evidence for the symmetric two-end model for FGFR dimerization. *Mol Cell Biol* **25**, 671-684.

**Ihermann-Hella, A., Lume, M., Miinalainen, I. J., Pirttiniemi, A., Gui, Y., Peranen, J., Charron, J., Saarma, M., Costantini, F. and Kuure, S.** (2014). Mitogen-activated protein kinase (MAPK) pathway regulates branching by remodeling epithelial cell adhesion. *PLoS Genet* **10**, e1004193.

**Ikonomou, L., Herriges, M. J., Lewandowski, S. L., Marsland, R., 3rd, Villacorta-Martin, C., Caballero, I. S., Frank, D. B., Sanghrajka, R. M., Dame, K., Kandula, M. M., et al.** (2020). The in vivo genetic program of murine primordial lung epithelial progenitors. *Nat Commun* **11**, 635.

**Ingham, P. W. and McMahon, A. P.** (2001). Hedgehog signaling in animal development: paradigms and principles. *Genes Dev* **15**, 3059-3087.

**Ioannides, A. S., Massa, V., Ferraro, E., Cecconi, F., Spitz, L., Henderson, D. J. and Copp, A. J.** (2010). Foregut separation and tracheo-oesophageal malformations: the role of tracheal outgrowth, dorso-ventral patterning and programmed cell death. *Dev Biol* **337**, 351-362.

**Jesudason, E. C., Smith, N. P., Connell, M. G., Spiller, D. G., White, M. R., Fernig, D. G. and Losty, P. D.** (2006). Peristalsis of airway smooth muscle is developmentally regulated and uncoupled from hypoplastic lung growth. *Am J Physiol Lung Cell Mol Physiol* **291**, L559-565.

**Jones, M. R., Dilai, S., Lingampally, A., Chao, C. M., Danopoulos, S., Carraro, G., Mukhametshina, R., Wilhelm, J., Baumgart-Vogt, E., Al Alam, D., et al.** (2018). A Comprehensive Analysis of Fibroblast Growth Factor Receptor 2b Signaling on Epithelial Tip Progenitor Cells During Early Mouse Lung Branching Morphogenesis. *Frontiers in genetics* **9**, 746.

**Kadzik, R. S., Cohen, E. D., Morley, M. P., Stewart, K. M., Lu, M. M. and Morrisey, E. E.** (2014). Wnt ligand/Frizzled 2 receptor signaling regulates tube shape and branch-point formation in the lung through control of epithelial cell shape. *Proceedings of the National Academy of Sciences of the United States of America* **111**, 12444-12449.

**Kadzik, R. S. and Morrisey, E. E.** (2012). Directing lung endoderm differentiation in pluripotent stem cells. *Cell Stem Cell* **10**, 355-361.

**Kim, H. Y., Pang, M. F., Varner, V. D., Kojima, L., Miller, E., Radisky, D. C. and Nelson, C. M.** (2015). Localized Smooth Muscle Differentiation Is Essential for Epithelial Bifurcation during Branching Morphogenesis of the Mammalian Lung. *Dev Cell* **34**, 719-726.

**Kim, H. Y., Varner, V. D. and Nelson, C. M.** (2013). Apical constriction initiates new bud formation during monopodial branching of the embryonic chicken lung. *Development* **140**, 3146-3155.

**Kimura, S., Hara, Y., Pineau, T., Fernandez-Salguero, P., Fox, C. H., Ward, J. M. and Gonzalez, F. J.** (1996). The T/ebp null mouse: thyroid-specific enhancer-binding protein is essential for the organogenesis of the thyroid, lung, ventral forebrain, and pituitary. *Genes Dev* **10**, 60-69.

**Kishimoto, K., Furukawa, K. T., Luz-Madrigal, A., Yamaoka, A., Matsuoka, C., Habu, M., Alev, C., Zorn, A. M. and Morimoto, M.** (2020). Bidirectional Wnt signaling between endoderm and mesoderm confers tracheal identity in mouse and human cells. *Nat Commun* **11**, 4159.

**Kishimoto, K., Tamura, M., Nishita, M., Minami, Y., Yamaoka, A., Abe, T., Shigeta, M. and Morimoto, M.** (2018). Synchronized mesenchymal cell polarization and differentiation shape the formation of the murine trachea and esophagus. *Nat Commun* **9**, 2816.

**Kurics, T., Menshykau, D. and Iber, D.** (2014). Feedback, receptor clustering, and receptor restriction to single cells yield large Turing spaces for ligand-receptor-based Turing models. *Phys Rev E Stat Nonlin Soft Matter Phys* **90**, 022716.

**Kuwahara, A., Lewis, A. E., Coombes, C., Leung, F. S., Percharde, M. and Bush, J. O.** (2020). Delineating the early transcriptional specification of the mammalian trachea and esophagus. *eLife* **9**.

**Lawson, K. A.** (1983). Stage specificity in the mesenchyme requirement of rodent lung epithelium in vitro: a matter of growth control? *J Embryol Exp Morphol* **74**, 183-206.

**Lin, Y., Zhang, S., Tuukkanen, J., Peltoketo, H., Pihlajaniemi, T. and Vainio, S.** (2003). Patterning parameters associated with the branching of the ureteric bud regulated by epithelial-mesenchymal interactions. *Int J Dev Biol* **47**, 3-13.

**Litingtung, Y., Lei, L., Westphal, H. and Chiang, C.** (1998). Sonic hedgehog is essential to foregut development. *Nat Genet* **20**, 58-61.

**Liu, Y., Martinez, L., Ebine, K. and Abe, M. K.** (2008). Role for mitogen-activated protein kinase p38 alpha in lung epithelial branching morphogenesis. *Dev Biol* **314**, 224-235.

**Liu, Y., Stein, E., Oliver, T., Li, Y., Brunken, W. J., Koch, M., Tessier-Lavigne, M. and Hogan, B. L.** (2004). Novel role for Netrins in regulating epithelial behavior during lung branching morphogenesis. *Curr Biol* **14**, 897-905.

**Lubkin, S. R. and Murray, J. D.** (1995). A mechanism for early branching in lung morphogenesis. *J Math Biol* **34**, 77-94.

**Madl, P., Hofmann, W., Oldham, M. J. and Asgharian, B.** (2010). Stochastic morphometric model of the BALB/c mouse lung. *Anat Rec (Hoboken)* **293**, 1766-1775.

**Mailleux, A. A., Kelly, R., Veltmaat, J. M., De Langhe, S. P., Zaffran, S., Thiery, J. P. and Bellusci, S.** (2005). Fgf10 expression identifies parabronchial smooth muscle cell progenitors and is required for their entry into the smooth muscle cell lineage. *Development* **132**, 2157-2166.

**Makarenkova, H. P., Hoffman, M. P., Beenken, A., Eliseenkova, A. V., Meech, R., Tsau, C., Patel, V. N., Lang, R. A. and Mohammadi, M.** (2009). Differential interactions of FGFs with heparan sulfate control gradient formation and branching morphogenesis. *Science signaling* **2**, ra55.

**Martin, A. C. and Goldstein, B.** (2014). Apical constriction: themes and variations on a cellular mechanism driving morphogenesis. *Development* **141**, 1987-1998.

**Mauroy, B., Filoche, M., Weibel, E. R. and Sapoval, B.** (2004). An optimal bronchial tree may be dangerous. *Nature* **427**, 633-636.

**Meinhardt, H.** (1976). Morphogenesis of lines and nets. *Differentiation* **6**, 117-123.

**Menshykau, D., Blanc, P., Unal, E., Sapin, V. and Iber, D.** (2014). An interplay of geometry and signaling enables robust lung branching morphogenesis. *Development* **141**, 4526-4536.

**Menshykau, D., Kraemer, C. and Iber, D.** (2012). Branch mode selection during early lung development. *PLoS Comput Biol* **8**, e1002377.

**Menshykau, D., Michos, O., Lang, C., Conrad, L., McMahon, A. P. and Iber, D.** (2019). Image-based modeling of kidney branching morphogenesis reveals GDNF-RET based Turing-type mechanism and pattern-modulating WNT11 feedback. *Nat Commun* **10**, 239.

**Metzger, R. J., Klein, O. D., Martin, G. R. and Krasnow, M. A.** (2008). The branching programme of mouse lung development. *Nature* **453**, 745-750.

**Miettinen, P. J., Warburton, D., Bu, D., Zhao, J. S., Berger, J. E., Minoo, P., Koivisto, T., Allen, L., Dobbs, L., Werb, Z., et al.** (1997). Impaired lung branching morphogenesis in the absence of functional EGF receptor. *Dev Biol* **186**, 224-236.

**Miller, L. A., Wert, S. E., Clark, J. C., Xu, Y., Perl, A. K. and Whitsett, J. A.** (2004). Role of Sonic hedgehog in patterning of tracheal-bronchial cartilage and the peripheral lung. *Dev Dyn* **231**, 57-71.

**Min, H., Danilenko, D. M., Scully, S. A., Bolon, B., Ring, B. D., Tarpley, J. E., DeRose, M. and Simonet, W. S.** (1998). Fgf-10 is required for both limb and lung development and

exhibits striking functional similarity to Drosophila branchless. *Genes Dev* **12**, 3156-3161.

**Minoo, P., Su, G., Drum, H., Bringas, P. and Kimura, S.** (1999). Defects in tracheoesophageal and lung morphogenesis in Nkx2.1(-/-) mouse embryos. *Dev Biol* **209**, 60-71.

**Miura, T. and Shiota, K.** (2002). Depletion of FGF acts as a lateral inhibitory factor in lung branching morphogenesis in vitro. *Mechanisms of Development* **116**, 29-38.

**Molladavoodi, S., Robichaud, M., Wulff, D. and Gorbet, M.** (2017). Corneal epithelial cells exposed to shear stress show altered cytoskeleton and migratory behaviour. *PLoS One* **12**, e0178981.

**Mollard, R. and Dziadek, M.** (1998). A correlation between epithelial proliferation rates, basement membrane component localization patterns, and morphogenetic potential in the embryonic mouse lung. *American journal of respiratory cell and molecular biology* **19**, 71-82.

**Moore, K. A., Polte, T., Huang, S., Shi, B., Alsberg, E., Sunday, M. E. and Ingber, D. E.** (2005). Control of basement membrane remodeling and epithelial branching morphogenesis in embryonic lung by Rho and cytoskeletal tension. *Dev Dyn* **232**, 268-281.

**Moore, M. W., Klein, R. D., Fariñas, I., Sauer, H., Armanini, M., Phillips, H., Reichardt, L. F., Ryan, A. M., Carver-Moore, K. and Rosenthal, A.** (1996). Renal and neuronal abnormalities in mice lacking GDNF. *Nature* **382**, 76-79.

**Motoyama, J., Liu, J., Mo, R., Ding, Q., Post, M. and Hui, C. C.** (1998). Essential function of Gli2 and Gli3 in the formation of lung, trachea and oesophagus. *Nat Genet* **20**, 54-57.

**Murray, J.** (1993). *Mathematical Biology*.

**Murray, J. D.** (1982). Parameter space for turing instability in reaction diffusion mechanisms: a comparison of models. *J Theor Biol* **98**, 143-163.

**Nelson, C. M., Gleghorn, J. P., Pang, M. F., Jaslove, J. M., Goodwin, K., Varner, V. D., Miller, E., Radisky, D. C. and Stone, H. A.** (2017). Microfluidic chest cavities reveal that transmural pressure controls the rate of lung development. *Development* **144**, 4328-4335.

**Nelson, T. R., West, B. J. and Goldberger, A. L.** (1990). The fractal lung: universal and species-related scaling patterns. *Experientia* **46**, 251-254.

**Nguyen, N. M. and Senior, R. M.** (2006). Laminin isoforms and lung development: all isoforms are not equal. *Dev Biol* **294**, 271-279.

**Nogawa, H., Morita, K. and Cardoso, W. V.** (1998). Bud formation precedes the appearance of differential cell proliferation during branching morphogenesis of mouse lung epithelium in vitro. *Dev Dyn* **213**, 228-235.

**Ohtsuka, N., Urase, K., Momoi, T. and Nogawa, H.** (2001). Induction of bud formation of embryonic mouse tracheal epithelium by fibroblast growth factor plus transferrin in mesenchyme-free culture. *Dev Dyn* **222**, 263-272.

**Onodera, T., Sakai, T., Hsu, J. C., Matsumoto, K., Chiorini, J. A. and Yamada, K. M.** (2010). Btbd7 regulates epithelial cell dynamics and branching morphogenesis. *Science* **329**, 562-565.

**Paine, I., Chauviere, A., Landua, J., Sreekumar, A., Cristini, V., Rosen, J. and Lewis, M. T.** (2016). A Geometrically-Constrained Mathematical Model of Mammary Gland Ductal Elongation Reveals Novel Cellular Dynamics within the Terminal End Bud. *PLoS Comput Biol* **12**, e1004839.

**Park, J., Zhang, J. J., Moro, A., Kushida, M., Wegner, M. and Kim, P. C.** (2010). Regulation of Sox9 by Sonic Hedgehog (Shh) is essential for patterning and formation of tracheal cartilage. *Dev Dyn* **239**, 514-526.

**Park, W. Y., Miranda, B., Lebeche, D., Hashimoto, G. and Cardoso, W. V.** (1998). FGF-10 is a chemotactic factor for distal epithelial buds during lung development. *Dev Biol* **201**, 125-134.

**Pepicelli, C. V., Kispert, A., Rowitch, D. H. and McMahon, A. P.** (1997). GDNF induces branching and increased cell proliferation in the ureter of the mouse. *Dev Biol* **192**, 193-198.

**Pepicelli, C. V., Lewis, P. M. and McMahon, A. P.** (1998). Sonic hedgehog regulates branching morphogenesis in the mammalian lung. *Curr Biol* **8**, 1083-1086.

**Peters, K., Werner, S., Liao, X., Wert, S., Whitsett, J. and Williams, L.** (1994). Targeted expression of a dominant negative FGF receptor blocks branching morphogenesis and epithelial differentiation of the mouse lung. *EMBO J* **13**, 3296-3301.

**Pichel, J. G., Shen, L., Sheng, H. Z., Granholm, A. C., Drago, J., Grinberg, A., Lee, E. J., Huang, S. P., Saarma, M., Hoffer, B. J., et al.** (1996). Defects in enteric innervation and kidney development in mice lacking GDNF. *Nature* **382**, 73-76.

**Que, J., Choi, M., Ziel, J. W., Klingensmith, J. and Hogan, B. L.** (2006). Morphogenesis of the trachea and esophagus: current players and new roles for noggin and Bmps. *Differentiation* **74**, 422-437.

**Que, J., Luo, X., Schwartz, R. J. and Hogan, B. L.** (2009). Multiple roles for Sox2 in the developing and adult mouse trachea. *Development* **136**, 1899-1907.

**Que, J., Okubo, T., Goldenring, J. R., Nam, K. T., Kurotani, R., Morrisey, E. E., Taranova, O., Pevny, L. H. and Hogan, B. L.** (2007). Multiple dose-dependent roles for Sox2 in the patterning and differentiation of anterior foregut endoderm. *Development* **134**, 2521-2531.

**Relan, N. K. and Schuger, L.** (1999). Basement membranes in development. *Pediatric and developmental pathology : the official journal of the Society for Pediatric Pathology and the Paediatric Pathology Society* **2**, 103-118.

**Resnick, A. and Hopfer, U.** (2007). Force-response considerations in ciliary mechanosensation. *Biophys J* **93**, 1380-1390.

**Rockich, B. E., Hrycaj, S. M., Shih, H. P., Nagy, M. S., Ferguson, M. A., Kopp, J. L., Sander, M., Wellik, D. M. and Spence, J. R.** (2013). Sox9 plays multiple roles in the lung epithelium during branching morphogenesis. *Proc Natl Acad Sci U S A* **110**, E4456-4464.

**Rossi, J. M., Dunn, N. R., Hogan, B. L. and Zaret, K. S.** (2001). Distinct mesodermal signals, including BMPs from the septum transversum mesenchyme, are required in combination for hepatogenesis from the endoderm. *Genes Dev* **15**, 1998-2009.

**Runser, S., Conrad, L., Gómez, H., Lang, C., Dumond, M., Sapala, A., Kramps, L., Michos, O., Vetter, R. and Iber, D.** (2020). The Biomechanical Basis of Biased Epithelial Tube Elongation.

**Rutledge, E. A., Parvez, R. K., Short, K. M., Smyth, I. M. and McMahon, A. P.** (2019). Morphogenesis of the kidney and lung requires branch-tip directed activity of the Adamts18 metalloprotease. *Dev Biol* **454**, 156-169.

**Saburi, S., Hester, I., Fischer, E., Pontoglio, M., Eremina, V., Gessler, M., Quaggin, S. E., Harrison, R., Mount, R. and McNeill, H.** (2008). Loss of Fat4 disrupts PCP signaling

and oriented cell division and leads to cystic kidney disease. *Nat Genet* **40**, 1010-1015.

**Sakai, T., Larsen, M. and Yamada, K. M.** (2003). Fibronectin requirement in branching morphogenesis. *Nature* **423**, 876-881.

**Sala, F. G., Del Moral, P. M., Tiozzo, C., Alam, D. A., Warburton, D., Grikscheit, T., Veltmaat, J. M. and Bellusci, S.** (2011). FGF10 controls the patterning of the tracheal cartilage rings via Shh. *Development* **138**, 273-282.

**Schittny, J. C.** (2017). Development of the lung. *Cell Tissue Res* **367**, 427-444.

**Schuger, L., O'Shea, S., Rheinheimer, J. and Varani, J.** (1990). Laminin in lung development: effects of anti-laminin antibody in murine lung morphogenesis. *Dev Biol* **137**, 26-32.

**Schuger, L., Skubitz, A. P., Zhang, J., Sorokin, L. and He, L.** (1997). Laminin alpha1 chain synthesis in the mouse developing lung: requirement for epithelial-mesenchymal contact and possible role in bronchial smooth muscle development. *J Cell Biol* **139**, 553-562.

**Sekine, K., Ohuchi, H., Fujiwara, M., Yamasaki, M., Yoshizawa, T., Sato, T., Yagishita, N., Matsui, D., Koga, Y., Itoh, N., et al.** (1999). Fgf10 is essential for limb and lung formation. *Nat Genet* **21**, 138-141.

**Serls, A. E., Doherty, S., Parvatiyar, P., Wells, J. M. and Deutsch, G. H.** (2005). Different thresholds of fibroblast growth factors pattern the ventral foregut into liver and lung. *Development* **132**, 35-47.

**Shakya, R., Jho, E.-h., Kotka, P., Wu, Z., Kholodilov, N., Burke, R., D'Agati, V. and Costantini, F.** (2005). The role of GDNF in patterning the excretory system. *Dev Biol* **283**, 70-84.

**Short, K., Hodson, M. and Smyth, I.** (2013). Spatial mapping and quantification of developmental branching morphogenesis. *Development* **140**, 471-478.

**Smith, B. M., Traboulsi, H., Austin, J. H. M., Manichaikul, A., Hoffman, E. A., Bleecker, E. R., Cardoso, W. V., Cooper, C., Couper, D. J., Dashnaw, S. M., et al.** (2018). Human airway branch variation and chronic obstructive pulmonary disease. *Proc Natl Acad Sci U S A* **115**, E974-E981.

**Snowball, J., Ambalavanan, M., Whitsett, J. and Sinner, D.** (2015). "Endodermal Wnt signaling is required for tracheal cartilage formation". *Developmental Biology* **405**, 56-70.

**Spooner, B. S. and Wessells, N. K.** (1970). Mammalian Lung Development - Interactions in Primordium Formation and Bronchial Morphogenesis. *Journal of Experimental Zoology* **175**, 445-&.

**Stopka, A., Kokic, M. and Iber, D.** (2019). Cell-based simulations of biased epithelial lung growth. *Phys Biol* **17**, 016006.

**Tang, N., Marshall, W. F., McMahon, M., Metzger, R. J. and Martin, G. R.** (2011). Control of mitotic spindle angle by the RAS-regulated ERK1/2 pathway determines lung tube shape. *Science* **333**, 342-345.

**Tang, Z., Hu, Y., Wang, Z., Jiang, K., Zhan, C., Marshall, W. F. and Tang, N.** (2018). Mechanical Forces Program the Orientation of Cell Division during Airway Tube Morphogenesis. *Dev Cell*.

**Tape, C. J., Ling, S., Dimitriadi, M., McMahon, K. M., Worboys, J. D., Leong, H. S., Norrie, I. C., Miller, C. J., Poulogiannis, G., Lauffenburger, D. A., et al.** (2016). Oncogenic KRAS Regulates Tumor Cell Signaling via Stromal Reciprocation. *Cell* **165**, 910-920.

**Turcatel, G., Rubin, N., Menke, D. B., Martin, G., Shi, W. and Warburton, D.** (2013). Lung mesenchymal expression of Sox9 plays a critical role in tracheal development. *BMC biology* **11**, 117.

**Turing, A. M.** (1952). The chemical basis of morphogenesis. *Phil. Trans. Roy. Soc. Lond* **B237**, 37-72.

**Unbekandt, M., del Moral, P. M., Sala, F. G., Bellusci, S., Warburton, D. and Fleury, V.** (2008). Tracheal occlusion increases the rate of epithelial branching of embryonic mouse lung via the FGF10-FGFR2b-Sprouty2 pathway. *Mech Dev* **125**, 314-324.

**Ustiyan, V., Zhang, Y., Perl, A. K., Whitsett, J. A., Kalin, T. V. and Kalinichenko, V. V.** (2016). beta-catenin and Kras/Foxm1 signaling pathway are critical to restrict Sox9 in basal cells during pulmonary branching morphogenesis. *Dev Dyn* **245**, 590-604.

**Usui, H., Shibayama, M., Ohbayashi, N., Konishi, M., Takada, S. and Itoh, N.** (2004). Fgf18 is required for embryonic lung alveolar development. *Biochem Biophys Res Commun* **322**, 887-892.

**Varner, V. D., Gleghorn, J. P., Miller, E., Radisky, D. C. and Nelson, C. M.** (2015). Mechanically patterning the embryonic airway epithelium. *Proc Natl Acad Sci U S A* **112**, 9230-9235.

**Volckaert, T., Campbell, A., Dill, E., Li, C., Minoo, P. and De Langhe, S.** (2013). Localized Fgf10 expression is not required for lung branching morphogenesis but prevents differentiation of epithelial progenitors. *Development* **140**, 3731-3742.

**Volckaert, T. and De Langhe, S. P.** (2015). Wnt and FGF mediated epithelial-mesenchymal crosstalk during lung development. *Dev Dyn* **244**, 342-366.

**Wan, X., Li, Z. and Lubkin, S. R.** (2008). Mechanics of mesenchymal contribution to clefting force in branching morphogenesis. *Biomech Model Mechan* **7**, 417-426.

**Warburton, D., Bellusci, S., Del Moral, P. M., Kaartinen, V., Lee, M., Tefft, D. and Shi, W.** (2003). Growth factor signaling in lung morphogenetic centers: automaticity, stereotypy and symmetry. *Respir Res* **4**, 5.

**Weaver, M., Dunn, N. R. and Hogan, B. L.** (2000). Bmp4 and Fgf10 play opposing roles during lung bud morphogenesis. *Development* **127**, 2695-2704.

**Weibel, E. R.** (1963). Geometry and Dimensions of Airways of Conductive and Transitory Zones. In *Morphometry of the Human Lung*, pp. 110-135. Berlin, Heidelberg: Springer Berlin Heidelberg.

**Weibel, E. R. and Gomez, D. M.** (1962). Architecture of the human lung. Use of quantitative methods establishes fundamental relations between size and number of lung structures. *Science* **137**, 577-585.

**Weinbaum, S., Duan, Y., Thi, M. M. and You, L.** (2011). An Integrative Review of Mechanotransduction in Endothelial, Epithelial (Renal) and Dendritic Cells (Osteocytes). *Cell Mol Bioeng* **4**, 510-537.

**Wessells, N. K. and Cohen, J. H.** (1968). Effects of collagenase on developing epithelia in vitro: Lung, ureteric bud, and pancreas. *Developmental Biology* **18**, 294-309.

**West, B. J., Bhargava, V. and Goldberger, A. L.** (1986). Beyond the principle of similitude: renormalization in the bronchial tree. *J Appl Physiol* **60**, 1089-1097.

**White, A. C., Xu, J., Yin, Y., Smith, C., Schmid, G. and Ornitz, D. M.** (2006). FGF9 and SHH signaling coordinate lung growth and development through regulation of distinct mesenchymal domains. *Development* **133**, 1507-1517.

**Wilson, T. A.** (1967). Design of the bronchial tree. *Nature* **213**, 668-669.

**Xu, H., Sun, M. and Zhao, X.** (2017). Turing mechanism underlying a branching model for lung morphogenesis. *PLoS One* **12**, e0174946.

**Yates, L. L., Schnatwinkel, C., Murdoch, J. N., Bogani, D., Formstone, C. J., Townsend, S., Greenfield, A., Niswander, L. A. and Dean, C. H.** (2010). The PCP genes Celsr1 and Vangl2 are required for normal lung branching morphogenesis. *Hum Mol Genet* **19**, 2251-2267.

**Yoshida, T., Matsuda, M. and Hirashima, T.** (2020). Incoherent Feedforward Regulation via Sox9 and ERK Underpins Mouse Tracheal Cartilage Development. *Frontiers in cell and developmental biology* **8**.

**Young, R. E., Jones, M. K., Hines, E. A., Li, R., Luo, Y., Shi, W., Verheyden, J. M. and Sun, X.** (2020). Smooth Muscle Differentiation Is Essential for Airway Size, Tracheal Cartilage Segmentation, but Dispensable for Epithelial Branching. *Dev Cell*.

**Yuan, B., Li, C., Kimura, S., Engelhardt, R. T., Smith, B. R. and Minoo, P.** (2000). Inhibition of distal lung morphogenesis inNkx2.1(?/?) embryos. *Developmental Dynamics* **217**, 180-190.

**Zeng, X., Goetz, J. A., Suber, L. M., Scott, W. J., Jr., Schreiner, C. M. and Robbins, D. J.** (2001). A freely diffusible form of Sonic hedgehog mediates long-range signalling. *Nature* **411**, 716-720.

**Zhu, X. and Yang, H.** (2018). Turing Instability-Driven Biofabrication of Branching Tissue Structures: A Dynamic Simulation and Analysis Based on the Reaction(-)Diffusion Mechanism (dagger). *Micromachines (Basel)* **9**.

# Tables

**Table 1: Reagents that impact lung branching morphogenesis**

| Reagents | Target | Growth | Branching | Bud Shape | Reference |
|---|---|---|---|---|---|
| BDM | Myosin ATPase | ↓ | ↓ | dilated | (Moore et al., 2005) |
| CNF-1 | Rho family GTPases | ? ↓ | ↑ 2ng/ml ↓ 200 ng/ml | normal | (Moore et al., 2005) |
| collagenase | collagen | NA | NA | cyst-like shapes | (Miura and Shiota, 2002; Wessells and Cohen, 1968) |
| cytochalasin D | actin | ↓ | ↓ | dilated | (Moore et al., 2005) |
| fasudil, Y27632 | ROCK | ↓ | ↓ | dilated | (Kadzik et al., 2014; Moore et al., 2005) |
| FGF7, FGF9 | FGFR | ↑ | ↓ | dilated | (White et al., 2006) |
| GM60001 | MMP | ↑ 2.5 μM ↓ 20 μM | ↑ 2.5 μM ↓ 20 μM | narrower wider | (Gill et al., 2003; Gill et al., 2006) |
| Anti-laminin ab | laminin | ↓ | ↓ | dilated | (Schuger et al., 1990) |
| ML7, ML9 | MLCK | ↓ | ↓ | dilated | (Kadzik et al., 2014; Moore et al., 2005) |
| nifedipine | Calcium channel | ↓ | ↑ | wider | (Goodwin et al., 2019) |
| SU5402 | FGFR | ↓ | ↓ | narrrower | (Goodwin et al., 2019; Runser et al., 2020) |

# FIGURES

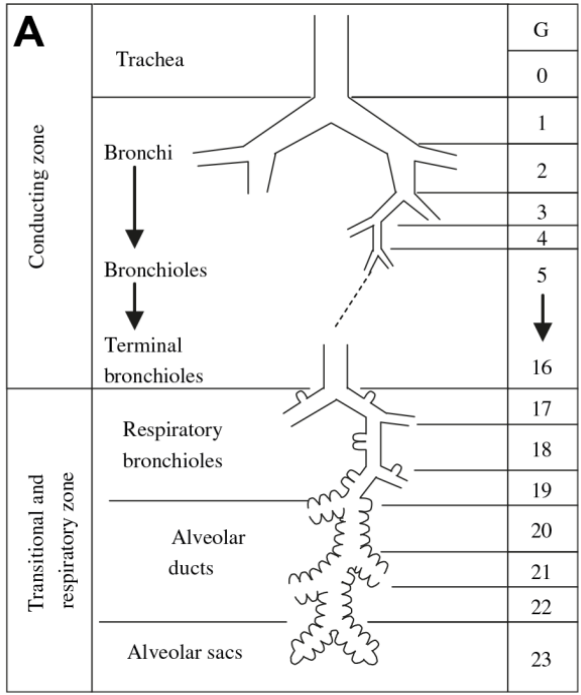
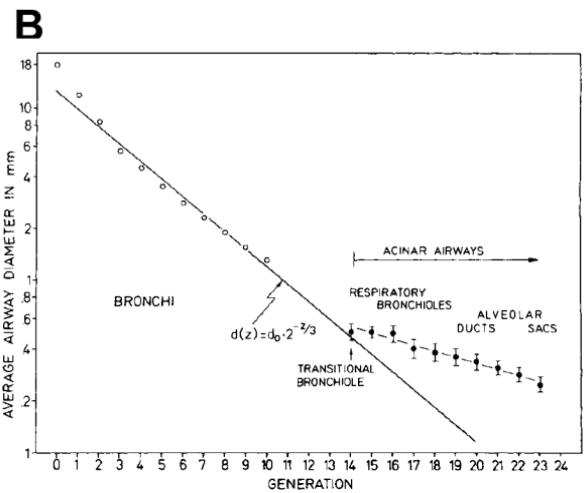
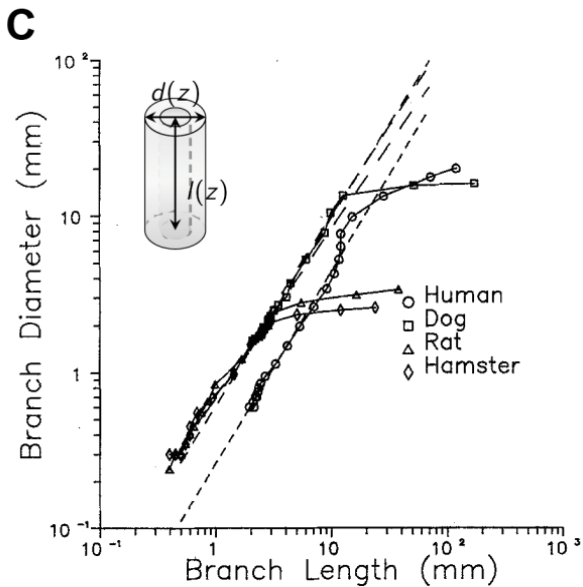

**Figure 1: The Adult Lung Tree**

**A)** The human respiratory airway can be divided into 24 various lung generations. Generation index 0–16 functions as the conducting zone, and generation index 17–23 functions as the transitional and respiratory zone. From (Ali, 2010), based on (Weibel, 1963).

**B)** The average airway diameter shrinks by a constant factor $2^{-1/3} \approx 0.8$, in each of the first 14 branch generations, whereas the intraacinar airways are reduced to a lesser degree. The graph was reproduced from (Haefeli-Bleuer and Weibel, 1988).

**C)** The ratio between the branch length and diameter is largely conserved between branch generations, but differs between species. Each point represents one bronchial order (z) where the trachea (z = 0) is the rightmost point. The graph was reproduced from (Nelson et al., 1990).

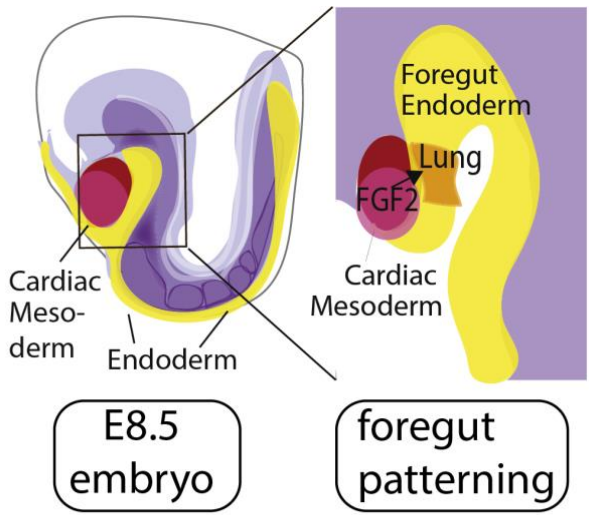
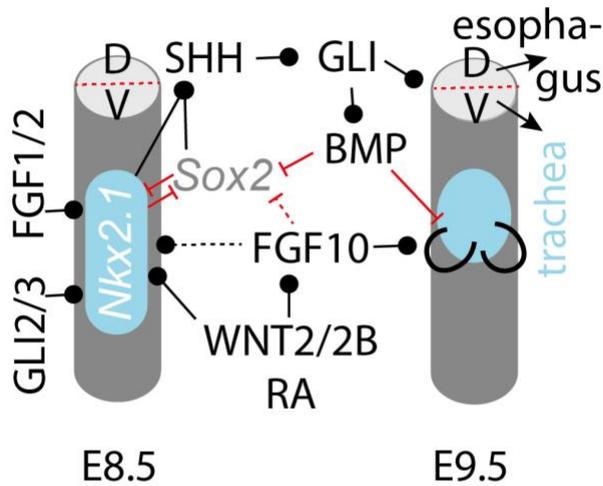
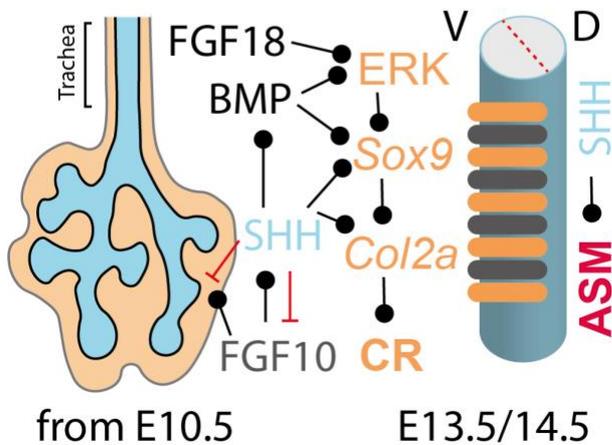

**Figure 2: Control of Lung Development**

The lung develops from the ventral anterior foregut endoderm, adjacent to the cardiac mesoderm.

**(top row)** At E8.5, the the definitive endoderm (yellow) is forming a foregut pocket. High levels of FGF1/2 from the cardiac mesoderm induce the lung field, marked by *Nkx2.1* expression.

**(middle and bottom rows)** Regulatory interactions that control the specification of the lung field (blue) on the ventral (V) side of the anterior foregut at E8.5, the formation of primary lung buds at E9.5, the separation of the trachea from the esophagus between E9.5 and E11, branching morphogenesis from E10.5, and the emergence of cartilage rings (CR) in the ventral and airway smooth muscle (ASM) in the dorsal (D) tracheal mesenchyme from E13.5. Black arrows indicate positive regulation, red arrows negative regulation. For details see text.

The image in the top row was reproduced from (Kadzik and Morrisey, 2012), the lung outline was modified from (Runser et al., 2020).

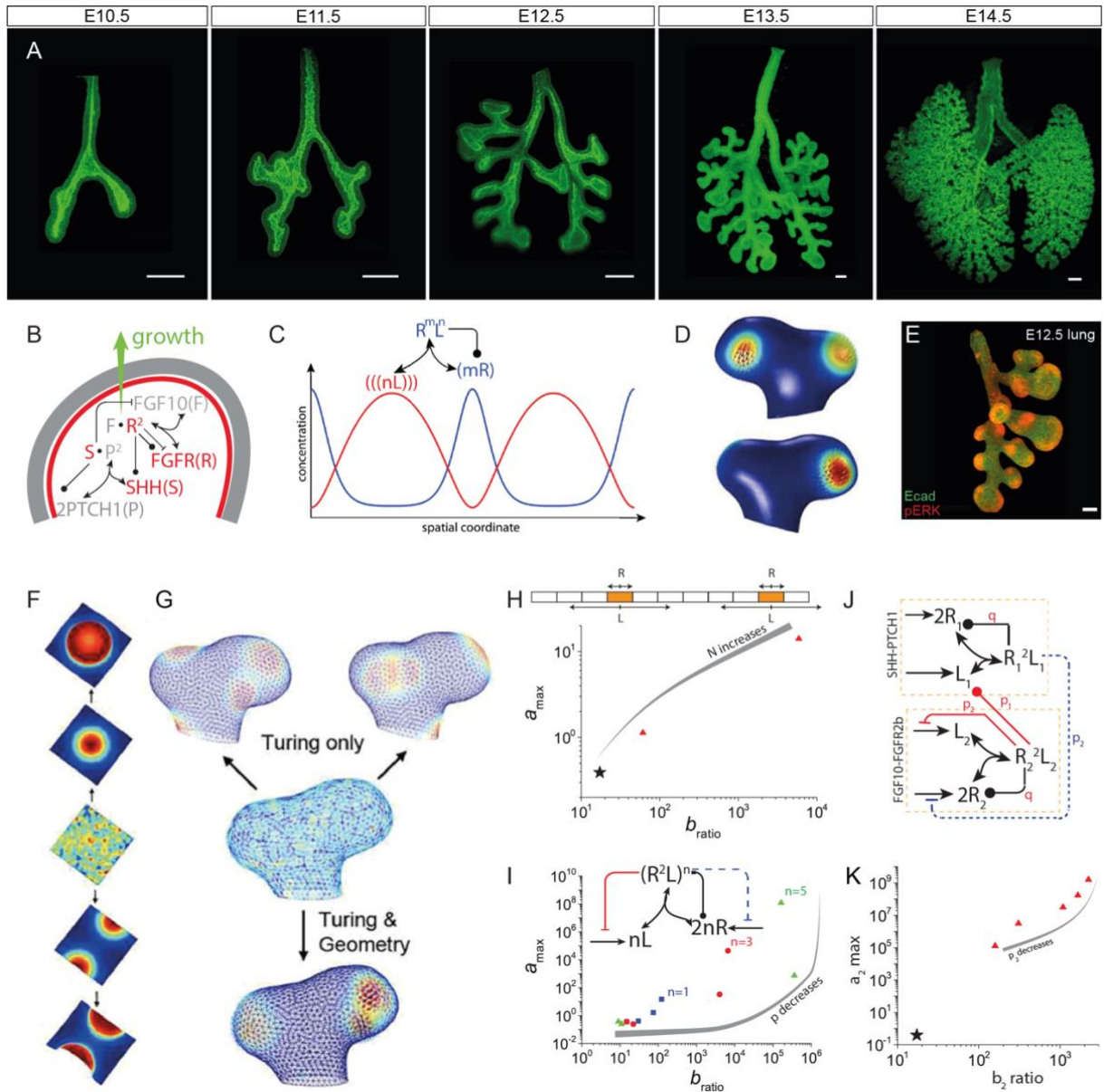

**Figure 3: Control of Lung Branching Morphogenesis by a ligand-receptor based Turing mechanism**

**A)** Developmental timeline of the mouse embryonic lung tree. Scale bars 200 μm.

**B)** The SHH-FGF10 network in the control of lung branching morphogenesis. *Fgf10* is expressed only in the mesenchyme (grey), while its receptor FGFR2b is restricted to the epithelium (red). FGF10 signalling induces bud formation (green arrow), and upregulates *Shh* expression and FGFR2b in the epithelium (red). SHH signalling represses *Fgf10* expression and upregulates the expression of its own receptor *Ptch1*.

**C)** Spatial patterns via a Turing mechanism can result from cooperative receptor-ligand interactions, where *m* receptors (R) and *n* ligand molecules (L) form an active complex that upregulates the receptor concentration by increasing its expression, limiting its turnover or similar. Importantly, the highest receptor and ligand concentrations are observed in different places.

**D)** The ligand-receptor-complex ($R^2L$) concentration pattern (solid colours) as predicted by the ligand-receptor-based Turing mechanism perfectly matches the 3D lung embryonic growth fields ass inferred from embryonic imaging data (arrows).

**E)** FGF signalling, as visualized by phosphorylated ERK (pERK, red) staining, is concentrated at the tips. Antibody staining for E-cadherin (green) marks the epithelium.

**F)** Noisy initial conditions can result in different patterns for the same parameter sets.

**G)** Ligand-receptor-based Turing mechanisms result in a wide range of different patterns for the same parameter set if solved only on the lung epithelium (top). Inclusion of the lung mesenchyme together with tissue-specific expression of ligand and receptor gives rise to a diffusion-based geometry effect that biases the Turing mechanism to a single pattern (bottom) in spite of noisy initial conditions (middle).

**H)** Diffusion of receptors is restricted to single cells, while ligand can diffuse over the entire domain. The range of receptor production rates, [0, $a_{max}$], and the fold-range of ligand production rates, $b_{ratio}$, that give rise to Turing patterns, the so-called Turing space, broadens as the domain of fixed size is split into more cells, *N*. Triangles show the results for $N = 10$ and $N = 100$ cells. The black star reports the Turing space for the standard model, $N = 1$.

**I)** Receptor clustering enlarges the Turing space. (Inset) The simulated network architecture. Clusters of $2n$ receptors $R$ interact with $n$ dimeric ligands $L$ to form a receptor-ligand complex $(R^2L)^n$ (black arrows, ↔). The receptor-ligand complex upregulates the presence of receptors (black interaction, ─•). In addition to these core interactions that can result in a Turing mechanism, there may be additional negative feedbacks on the ligand production (red solid arrow) and/or the receptor production (blue dashed arrow). Higher cooperativity, $n > 1$, as may result from larger receptor-ligand clusters, increases the size of the Turing space. The gray arrow indicates the direction, in which the feedback threshold, $p$, decreases.

**J)** Schematic representation of the regulatory network for lung branching morphogenesis in panel (B). ↔ indicates binding interactions, -| indicates inhibitory interactions, and ─• indicates up-regulating interactions.

**K)** The Turing space of the physiological model in panels B,I is huge and further increases as the feedback threshold, $p$, is lowered. The red triangles represent the Turing spaces for $p_1 = q = 0.1$ (positive feedback on ligand and receptor, respectively) and $p_2 = 0.01, 0.1, 1, 10, 100$ (negative feedback); the black star represents the size of the Turing space of the standard network (black part of networks). The gray arrow indicates the direction in which the feedback threshold, $p$, decreases.

Panels A, D were reproduced from (Runser et al., 2020), panels B, C, H-K from (Kurics et al., 2014), panels E-G from (Menshykau et al., 2014).

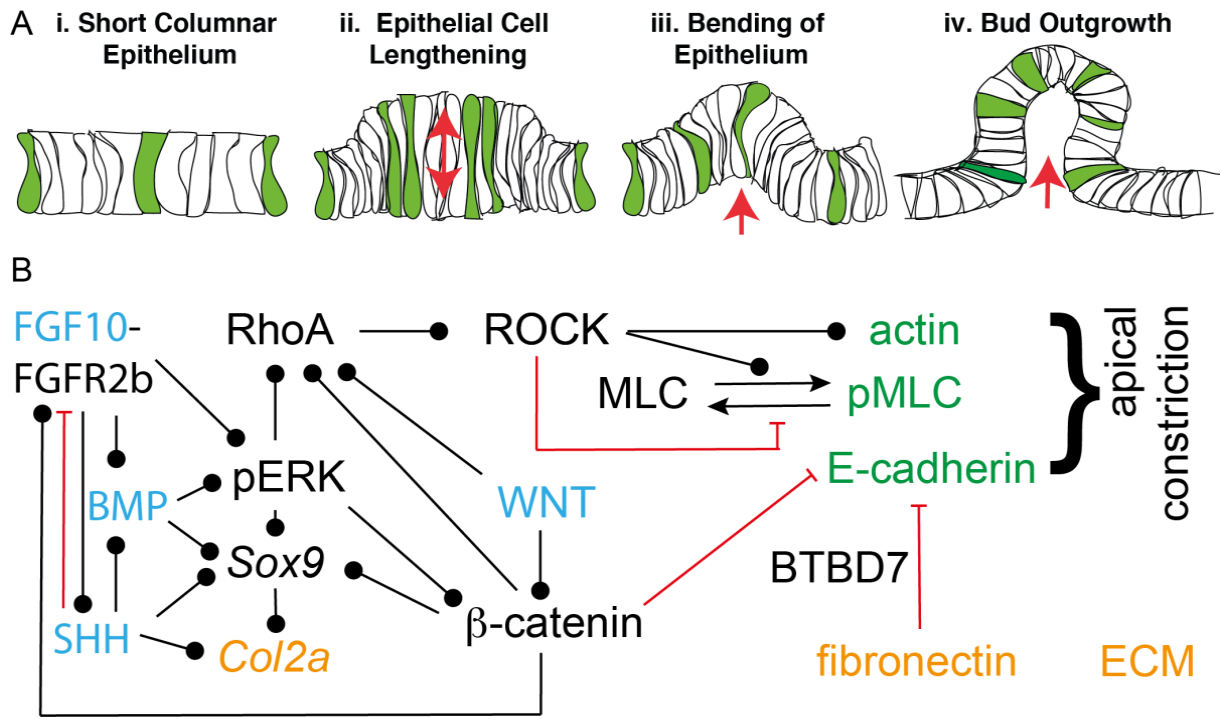

**Figure 4: Lung Bud Formation**

**A)** Diagram showing how the epithelial sheet that composes the developing lung airways deforms at sites of new branchpoint formation: Epithelial cells lengthen along the apical–basal axis, creating a bend in the epithelial sheet. This process results in the formation of new bud or branch point in the epithelial tube. As the new bud extends, the cells in the bud tip once again adopt a short columnar morphology. The picture and legend was reproduced from (Kadzik et al., 2014).

**B)** Simplified network of regulatory interactions that control cell mechanics in the lung epithelium. The morphogens FGF10, SHH, BMP and WNT are marked in blue, the ECM components in orange, and the components of the adhesion belt in green. Arrows also reflect indirect regulatory interactions. For details see text.

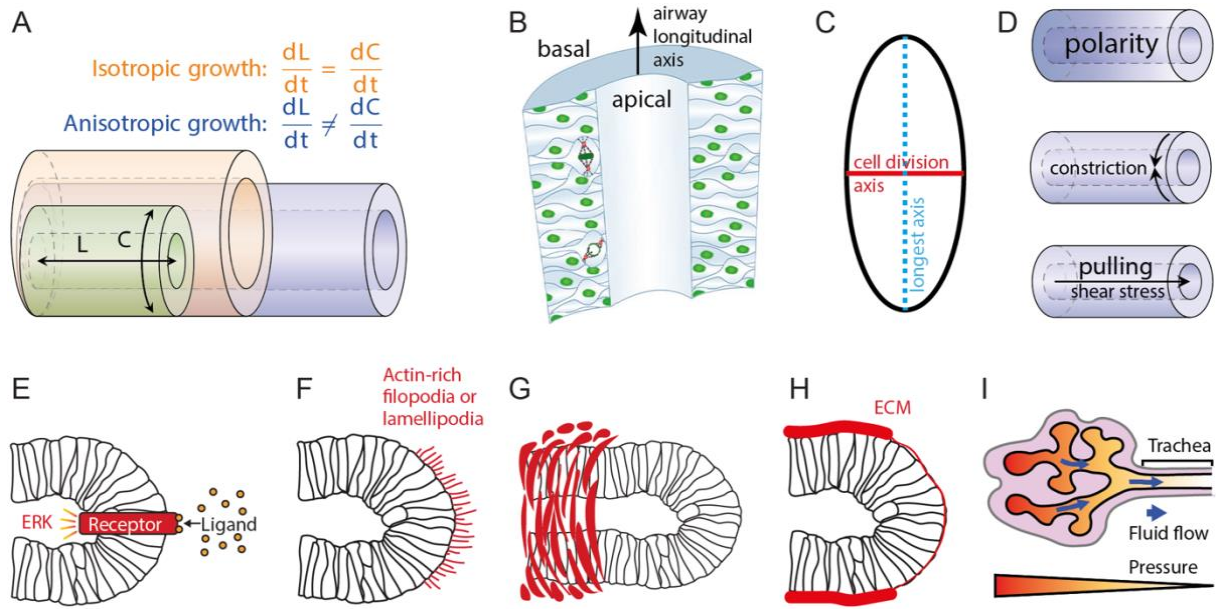

**Figure 5: Control of Lung Tube Elongation**

**A)** Schematic of isotropic and anisotropic tube expansion during development.

**B)** The cell shapes and mitotic angles are biased towards the direction of outgrowth.

**C)** Longest axis rule in the growing lung tubes: cells divide perpendicular (red line) to their longest axis (dashed blue line) when the cell aspect ratio at interphase is higher than 1.53 (Tang et al., 2018).

**D)** Potential drivers of elongating outgrowth: a polarity cue along the epithelial tubes, a constricting force from the mesenchyme or ECM, or a pulling force generated by the mesenchyme, the ECM, or via fluid flow-induced shear stress.

**E)** FGF10 signalling concentrates at the tips, but biased elongating outgrowth is observed also when mesenchyme-free lung explants are cultured with uniform FGF10, and when FGF receptor signalling is inhibited.

**F)** Actin-rich filopodia or lamellipodia could pull the epithelial tubes in direction of outgrowth, but are not observed in developing lung buds.

**G)** Constricting forces from the mesenchyme are not required for the biased elongating outgrowth of lung buds.

**H)** The ECM is thinner at the epithelial tips, and ECM turn-over affects the speed of outgrowth.

**I)** Fluid-flow induced shear stress could bias elongating outgrowth.

Panels A, E-I were reproduced with modifications from (Runser et al., 2020), panel B from (Tang et al., 2018).